%% file: revision3.tex
\documentclass[useAMS,usenatbib]{mnras}
\usepackage{graphicx}
\usepackage{amsmath}
\usepackage{float}

\title[Examining molecular clouds in the Galactic Centre region using X-ray reflection spectra simulations.]
  {Examining molecular clouds in the Galactic Centre region using X-ray reflection spectra simulations.}
\author[M. Walls et al.]
{M.~Walls,$^1$
  M.~Chernyakova,$^1$$^,$$^2$ R.~Terrier,$^3$ A. Goldwurm,$^3$$^,$$^4$ \\
  $^1$School of Physical Sciences, Dublin City University, Glasnevin, Dublin 9, Ireland\\
  $^2$Dublin Institute of Advanced Studies, 31 Fitzwilliam Place, Dublin 2, Ireland\\
  $^3$Astroparticule et Cosmologie, Universit´e Paris 7 Denis Diderot, 75205 Paris Cedex 13, France\\
  $^4$Service d’Astrophysique/IRFU/DRF CEA-Saclay, Bat. 709, 91191 Gif-sur-Yvette, France}
\date{Released 2016 Xxxxx XX}

\pagerange{\pageref{firstpage}--\pageref{lastpage}} \pubyear{2016}

\def\LaTeX{L\kern-.36em\raise.3ex\hbox{a}\kern-.15em
    T\kern-.1667em\lower.7ex\hbox{E}\kern-.125emX}

\begin{document}

\label{firstpage}

\maketitle

\begin{abstract}
In the centre of our galaxy lies a super-massive black hole, identified with the radio source Sagittarius $A^\star$. This black hole has an estimated mass of around 4 million solar masses. Although Sagittarius $A^\star$ is quite dim in terms of total radiated energy, having a luminosity that is a factor of $10^{10}$ lower than its Eddington luminosity, there is now compelling evidence that this source was far brighter in the past. Evidence derived from the detection of reflected X-ray emission from the giant molecular clouds in the galactic centre region. However, the interpretation of the reflected emission spectra cannot be done correctly without detailed modelling of the reflection process. Attempts to do so can lead to an incorrect interpretation of the data. In this paper we present the results of a Monte Carlo simulation code we developed in order to fully model the complex processes involved in the emerging reflection spectra. The simulated spectra can be compared to real data in order to derive model parameters and constrain the past activity of the black hole. In particular we apply our code to observations of Sgr B2, in order to constrain the position and density of the cloud and the incident luminosity of the central source. The results of the code have been adapted to be used in Xspec by a large community of astronomers.
\end{abstract}


\section{Introduction} \label{introduction}
In the centre of our Galaxy lies a super-massive black hole, identified with the bright radio source Sagittarius $A^\star$ (Sgr $A^\star$). The black hole has an estimated mass of ~4 million solar masses \citep{gillessen09}, and currently it is very faint with an X-Ray luminosity of only 10$^{33}$ - 10$^{34}$ erg/s \citep{baganoff03}, orders of magnitude dimmer than other Active Galactic Nuclei (AGN). During times of flaring the luminosity of Sgr $A^\star$ is known to increase by a factor of 100, but even this luminosity is still 10$^8$ lower than the Eddington luminosity for a black hole of this mass  \citep{baganoff01}. However it is possible that previously Sgr $A^\star$ was much brighter than it is now. The X-ray emission from the Central Molecular Zone (CMZ) was first interpreted as a reflection of a past flaring of Sgr $A^\star$ by \cite{sunyaev93, koyama96}.

Since then several studies have confirmed this hypothesis. Indeed for the past decade XMM-Newton has been regularly monitoring the the CMZ, along with other X-ray satellites, such as \textit{INTEGRAL}, \textit{Suzaku}, \textit{Chandra} and \textit{NuSTAR}. These observations have shown that the X-ray emission from some of the molecular clouds varies significantly with time \citep{muno07, inui09, ponti10, terrier10, clavel13, clavel14, zhang15}. For example the most massive cloud in the region, Sgr B2 has shown a relatively constant X-ray flux up to 2003, and since then has been showing a large flux decay, while another cloud, known as "The Bridge", has shown an increase in intensity. Some clouds in the region show constant X-ray emission, and others display hardly any X-ray emission at all. This non-uniform behaviour disfavours the interpretation of the X-ray emission as a result of accelerated particle interactions. The hypothesis that the illuminating source be Sgr $A^\star$ is supported by the large luminosity ( $> 10^{39}$ erg/s) needed to explain the observed flux of the Giant Molecular Clouds (GMC) \citep{ponti13}.

The morphological evidence for a reflection origin of the emission is also compelling. The overall symmetry around the source of the Fe $K_\alpha$ emission (observed from GMC with both positive and negative galactic longitudes) along with the fact that the emission from Sgr B2 is shifted toward the galactic centre by about ~2 arcmin when compared to the molecular mass distribution, suggesting the illuminating source is in the direction of the Galactic Centre (See \cite{murakami01, terrier10, ponti13} for an in-depth look at the work in this area to date).

Sgr B2 is a good candidate for observing the reflected X-rays of Sgr $A^\star$. Sgr B2 is the largest, most massive and dense cloud of the CMZ \citep{protheroe08}. Sgr B2 has been studied as an X-Ray Reflection Nebula (XRN) by several authors using ASCA, \textit{Chandra}, \textit{INTEGRAL}, \textit{XMM-Newton}, \textit{Suzaku} and \textit{NuSTAR} observations \citep{koyama96, murakami01, revnivtsev04, koyama07, terrier10, ponti13, zhang15} and the general interpretation is a reflection nebula generated by a Sgr $A^\star$ outburst occurring between 100 and 300 years past. However to understand the timing, durations and intensities of any outbursts from Sgr $A^\star$ that are being reflected by Sgr B2, it is necessary to know the position of Sgr B2 relative to Sgr $A^\star$ \citep{ponti13}. The latter is not yet well constrained. \cite{reid09} have measured the parallax of Sgr B2 and Sgr $A^\star$ with VLBI observations and deduced that the former should lie $\approx 120 $ pc in front of the latter assuming circular rotation. Conversely, in the \cite{molinari11} model, it lies far behind. The improved model of \cite{kruijssen15}, which takes into account orbital motion in the GC, gives a line of sight position of \textbf{$\approx$38 pc} in front of Sgr $A^\star$. 

Many analyses of the X-ray reflection phenomenon in the GC have relied on the X-ray spectrum only to extract information on the cloud thickness and location and deduce constraints on the illuminating source luminosity and spectrum. The spectrum of an X-ray nebula is often assumed to follow a simple absorbed power-law shape. In some cases, specific models of reflection spectra are used such as {\tt pexrav} \citep{ponti10} or {\tt MyTorus} \citep{zhang15, mori15}. However, we argue that these models do not reproduce precisely the spectral shape of an appropriate XRN, because they do not model the geometry of an isolated illuminated cloud. Specifically, the {\tt pexrav} model is a cold disk viewed from a given angle, is only valid in compton thick cases, does not re-produce the iron line, and does not provide a determination of the cloud column density $N_H$. Both \cite{zhang15} and \cite{mori15} have argued that the {\tt MyTorus} model is valid and applicable. However, there are some notable shortcomings of using the {\tt MyTorus} model, which are outlined in the appendices of both \cite{zhang15} and \cite{mori15}. Primarily {\tt MyTorus} only deals with uniform density, has a fixed iron abundance and there is also the geometry of {\tt MyTorus}, with scattering from the far side of the torus being an issue.

This issue can be critical when the spectral determination is used to constrain the geometry of the reflection, namely the position of the cloud with respect to the illuminating source. For example, some authors have used the fitted $\mathrm{N_H}$ of the absorbed power-law to estimate the column density of the cloud and hence its location along the line of sight using partial covering of the plasma emissions in the CMZ \citep{ryu09}. Others have used an estimate of the equivalent width (EW) of the 6.4 keV line to place the cloud along the line of sight \citep{capelli12}. Both approaches are sensitive on the determination of the underlying continuum level and require correct models of the scattered X-ray spectrum.

To address this issue it is necessary to produce a reliable model of reflected X-rays from an isolated cloud that can be properly applied to X-ray observations (i.e. through forward folding methods). Reflection spectra for Compton thin constant density spheres can be calculated analytically (See Appendix A). Indeed in section \ref{code_output} we present the results from these semi-analytical calculations. However these calculations are only dealing with Klein-Nishina scattering and are not taking into account binding modified multiple scattering (as such they are invalid in the Compton thick case), complex geometries or non-uniform densities. To this end we create a Monte Carlo (MC) code to simulate X-ray reflection spectra from molecular clouds and use it to constrain such properties of Sgr B2 as the dense central core column density $N_H$ , the photon index $\Gamma$ of the incident X-ray emission, the incident Sgr $A^\star$ luminosity and the position of Sgr B2. We show that the position of a molecular cloud has a strong influence both on the shape of the continuum and the relative strength of the 6.4 keV neutral iron line. Using an Xspec table model, we can compare model spectra to existing data and place constraints on the location of the Sgr B2 molecular cloud.

Other authors have developed Monte Carlo models of the reflection process in cold molecular material, for example the early work of \cite{sunyaev98, murakami01}. However these works were subject to large approximations such as a fixed photon index and isotropic scattering as well as fixed geometry (in particuler the viewing angle). The work of \cite{odaka11} is free from such approximations and is comparable to our own. Their work concentrated more on the reflection morphology, while we concentrate on the spectral features and produce the Xspec table model to allow for fitting to real observational data. There is also the work by \cite{molaro16} which concentrated on the effects of clumpiness in the cloud rather than on the spectral features.

This paper is organised as follows. In Section \ref{the_mc_code} we describe the details of our MC code, then in Section \ref{code_output} we discuss the dependence of the reflected component on the parameters of the model. In Section \ref{observations} we present the data obtained by \textit{Chandra}, \textit{XMM-Newton} and \textit{Integral} of Sgr B2, while in Section \ref{sgrb2} we apply the model to this data. Finally we compare our findings with previous work and present our conclusions in Section \ref{discussion}.

\section{The Monte Carlo Code} \label{the_mc_code}
\begin{figure}
\includegraphics[scale=0.4]{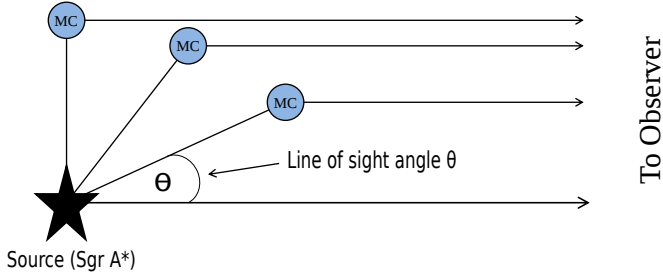}
 \caption{Geometry of the simulated system composed of an illuminating source (star) a reflecting cloud (MC) and an observer at infinity. The line of sight angle theta ($\theta$) is shown.}
  \label{fig:basic}
\end{figure}

We wish to constrain primarily the line of sight angle $\theta$ of XRN, see Figure \ref{fig:basic}. The code structure follows an approach quite similar to that of \cite{leahy93}. The source is assumed to have a power law spectrum, thus we simulate such an incident spectrum within an energy range of E$_{min}$ to E$_{max}$. The photons movement within a cold gas cloud and all of their interactions therein are simulated. The output spectrum is constructed from photons escaping the cloud in the direction of the observer. Within the cloud the photons may be absorbed, Rayleigh scatter or Compton scatter. They may also undergo fluorescence interactions with the K-shell electrons of atoms, for simplicity and clarity we neglect this process for all elements except iron, which has a significant chance of fluorescence.

In the simulation there is an absolute coordinate system in which the photons movement is allowed. The cloud position can be moved freely within the absolute coordinate system. But for our purposes the polar angle and distance from source to cloud were kept constant and the azimuthal angle was variable. Typically any single input spectra will consist of $\sim 10^{9}$ photons.

We model non-uniform density using three density profiles, namely $r^{-2}$, $e^{-r}$ and Gaussian. Unfortunately the limits of computation time mean that modelling a non-uniform density can only be approximate, the density is taken to be constant in steps of a certain size $\kappa$. The photon is thus propagated through small areas of constant density, with the density being re-calculated for each step. The smaller the step size the more accurate the result but the longer the computational time.

The source emits its photons sequentially following a power law distribution. The probability distribution function for photon production is given by P(E) = NE$^{-\Gamma}$, where $\Gamma$ is the photon index and N is the normalisation constant. For our purposes E$_{min}$ and E$_{max}$ are set to 1.0 and 400 keV respectively. 

The photons may either Rayleigh (coherent) or Compton (incoherent) scatter off hydrogen molecules and helium atoms in the cloud. The scattering contribution of heavier elements is negligible and they are ignored. It is important not to ignore the effect that scattering from bound electrons has on the cross sections, a lot of previous work in this area has treated the electrons as unbound and used pure Klein-Nishina scattering. As shown by \cite{molaro14} binding effects must be taken into account when considering scattering from molecular hydrogen. Thus the differential Rayleigh scattering cross section is taken to be the the Thomson differential cross section, modified by the atomic form factor, see equation (\ref{eq:ray+ff}).

\begin{equation}\label{eq:ray+ff}
\frac{d\sigma_R}{d\Omega} = \frac{d\sigma_{Th}}{d\Omega}\times F^{2}(x,Z) = \frac{r_e^2}{2}(1+cos^{2}\theta)\times F^{2}(x,Z),
\end{equation}

where $r_e = \frac{e^2}{m_e c^2}$ is the classical electron radius, $F^{2}(x,Z)$ is the atomic form factor, with Z being the atomic number of the absorber and $x = \frac{sin(\frac{\theta}{2})}{\lambda}$ is the momentum transfer variable with $\lambda$ being the wavelength of the photon.\\

Atomic form factors are only possible to calculate analytically for hydrogen, however \cite{hubbell75} calculated and tabulated the values for the heavier elements. We make use of these tables through {\tt xraylib}\footnote{https://github.com/tschoonj/xraylib} \citep{schoonjans76}. {\tt xraylib} provides these values through their easy access library. Because we are dealing with molecular hydrogen and not atomic, we make use of the \cite{sunyaev99} approximation and the Rayleigh scattering cross section is multiplied by a factor of two. The differential Compton scattering cross section is taken to be the Klein-Nishina differential scattering cross section modified by the incoherent scattering function, see equation (\ref{eq:com+isf}).

\begin{equation}\label{eq:com+isf}
\frac{d\sigma_C}{d\Omega} = \frac{d\sigma_{KN}}{d\Omega}\times S(x,Z),
\end{equation}

where $S(x,Z)$ is the incoherent scattering function. With Z being the atomic number of the absorber and $x = \frac{sin(\frac{\theta}{2})}{\lambda}$ is the momentum transfer variable with $\lambda$ being the wavelength. $\frac{d\sigma_{KN}}{d\Omega}$ is given by \ref{eq:kn}.

\begin{equation}\label{eq:kn}
\frac{d\sigma_{KN}}{d\Omega} = \frac{1}{2r_{e}^{2}}\;\frac{1+\cos^2{\theta}}{(1+2\epsilon\sin^2{}\frac{\theta}{2})^2}
\Bigg\{ 1 + \frac{4\epsilon^2\sin^4{\frac{\theta}{2}}}{[1+\cos^2{\theta}][1+2\epsilon\sin^2{}\frac{\theta}{2}]} \Bigg\},
\end{equation}

where $\epsilon = \frac{E_0}{m_e c^2}$, $r_e = \frac{e^2}{m_e c^2}$ is the classical electron radius and $\theta$ is the scattering angle.\\ 

If the photon is determined to have scattered it is given a new direction of propagation and in the case of Compton scattering it loses energy with the new energy being given by equation (\ref{eq:scat_en_loss}). 

\begin{equation}\label{eq:scat_en_loss}
E^{'} = E\frac{1}{1+\frac{E}{m_e c^2}(1-\cos{\theta})},
\end{equation}

where $\theta$ is the angle through which the photon has scattered.

The new direction of propagation relative to the photon propagation direction is found by means of rejection sampling the differential scattering cross sections, equations \ref{eq:ray+ff} and \ref{eq:com+isf}. In order to have reasonable computational times multiple scattering is limited to 35 scatters per photon, i.e. if a photon attempts to scatter for a 36th consecutive time before leaving the cloud, it will be terminated; it will not be scattered again, absorbed, re-emitted or observed. This limitation has a negligible effect for the density ranges we are interested in. Indeed multiple scattering beyond 10 scatters only has a noticeable effect in the most extreme of cases, that is very high density and very hard incident spectra.

The absorption cross sections are taken from the NIST XCOM database\footnote{http://www.nist.gov/pml/data/xcom/}. The cloud is considered to be cold (T $<$ 10$^6$K) thus not producing it's own X-rays. Solar abundances were used with values taken from \cite{lodders03}. If the photon is determined to have been absorbed there is a chance that it will be re-emitted by an iron atom. The probability that it was absorbed by iron is taken from the aforementioned abundances. The fluorescence yield is taken to be 0.34 \citep{bambynek72}. The probability of K$_\beta$ fluorescence relative to that of K$_\alpha$ fluorescence is taken to be 0.13 \citep{kaastra93}. K$_{\alpha1}$ and K$_{\alpha2}$ are treated as one given the low resolution of current instruments. The energy of a K$_\alpha$ fluorescence is taken to be 6.399 keV and that of K$_\beta$ is taken to be 7.085 keV\footnote{Although detailed iron line profiles are not covered here, future work will deal with it.}.

All photons that leave the cloud are recorded in bins of equal solid angle. The desired angular spectra is then extracted from this complete emergence spectral data. The photons are also binned by energy in logarithmically spaced energy bins. Of note is the fact we are only considering scattered spectra, i.e. only photons that scatter through very slight angles will be recorded in the 0 degree case. This makes the 0 degree case wholly un-physical. However, it is still useful in understanding the output spectra.

\section{Code output analysis} \label{code_output}
\subsection{Continuum}
\begin{figure}
\includegraphics[scale=0.7]{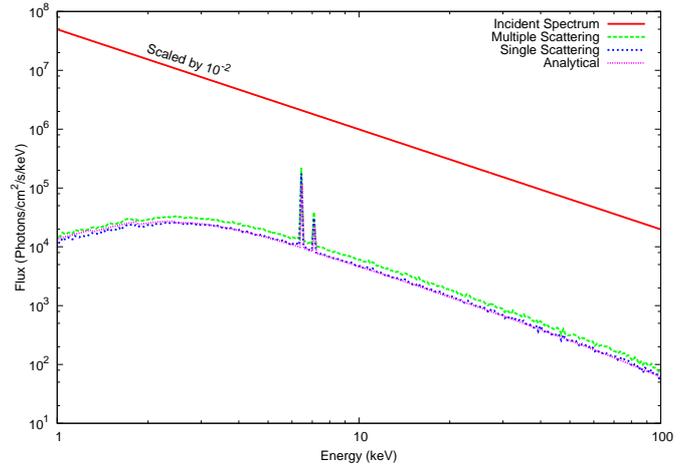}
\caption{Comparison of the spectra produced by the Monte Carlo code when considering only Klein-Nishina scattering in Single and Multiple scatter modes versus an analytically calculated spectrum for a 2 pc diameter uniform density sphere. Line of sight angle $\theta = 120^{\circ}$ and $N_H$ = $5\times10^{22}$ cm$^{-2}$. The incident spectrum has been downscaled by $10^{-2}$ to aid visualization.}
\label{fig:ana-v-mc-lownh}
\end{figure}

\begin{figure}
\includegraphics[scale=0.7]{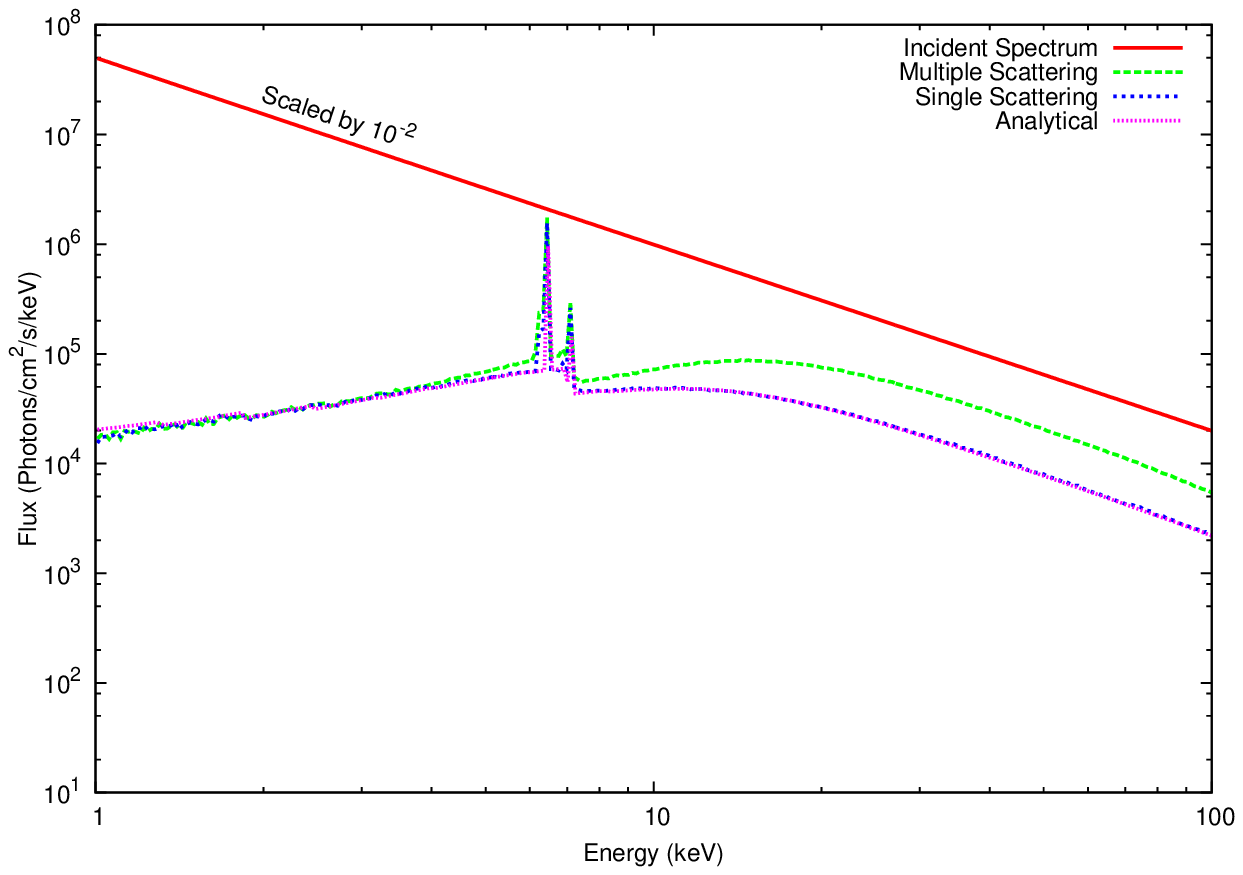}
 \caption{Comparison of the spectra produced by the Monte Carlo code when considering only Klein-Nishina scattering in Single and Multiple scatter modes versus an analytically calculated spectrum for a 2 pc diameter uniform density sphere. Line of sight angle $\theta = 120^{\circ}$ and $N_H$ = $5\times10^{24}$ cm$^{-2}$. The incident spectrum has been downscaled by $10^{-2}$ to aid visualization.}
\label{fig:ana-v-mc-hinh}
\end{figure}

As mentioned in section \ref{introduction}, the reflection spectra of a Compton thin, uniform density sphere can be calculated analytically (see Appendix A). We use this analytical spectrum to verify the MC code. When considering only Klein-Nishina scattering and in single scatter mode for a uniform density sphere the MC code is in good agreement with the analytical calculation, although there is a slight discrepancy below 2 keV. When we account for multiple scattering we begin to see a discrepancy between single and multiple scattering in the low $N_H$ case which becomes far more pronounced as $N_H$ increases, see Figures \ref{fig:ana-v-mc-lownh} and \ref{fig:ana-v-mc-hinh} where we see that the analytical and single scattering MC are in agreement and the multiple scattering is divergent. This underlines the importance of using the Monte Carlo technique.

\begin{figure}
\includegraphics[scale=0.7]{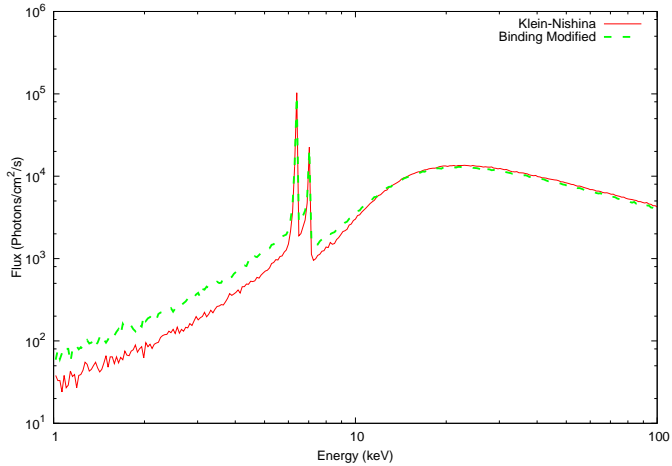}
 \caption{Comparison of the spectra produced by the Monte Carlo code using Klein-Nishina scattering versus binding modified Rayleigh and Compton scattering for a 2 pc diameter uniform density sphere. Line of sight angle $\theta = 30^{\circ}$ and an $N_H$ = $5\times10^{24}$ cm$^{-2}$.}
\label{fig:rc-vs-kn}
\end{figure}

Accounting for binding modified scattering, as discussed in section \ref{the_mc_code}, leads to further discrepancy particularly in the low energy regime. See Figure \ref{fig:rc-vs-kn} where the red spectrum is pure Klein-Nishina scattering on free electrons and the green is the binding modified Rayleigh + Compton scattering. This discrepancy underlines the importance of considering binding effects when considering scattering from molecular hydrogen.

\begin{figure}
\includegraphics[scale=0.7]{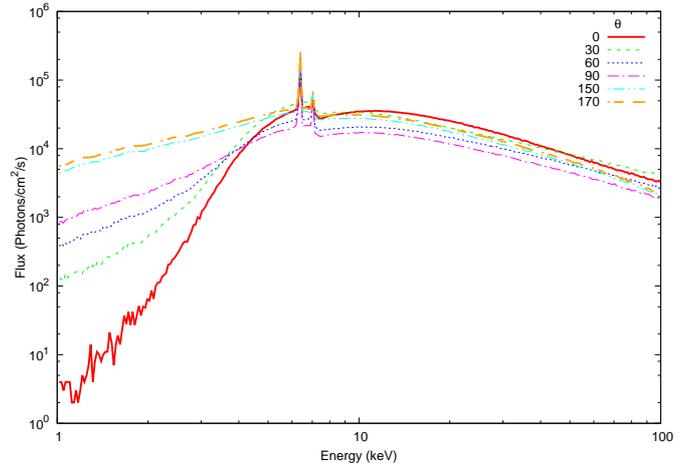}
 \caption{Spectra produced by the MC code for a 2 pc diameter, uniform density sphere with an $N_H$ of $6\times10^{23}$ cm$^{-2}$, and a photon index of 2.0. Which shows the changes in flux and continuum shape resulting from a changing line of sight angle.}
  \label{fig:changing_angle}
\end{figure}

The spectral shape and total flux of the scattered emission is very dependant on the relative position of the cloud, see Figure \ref{fig:changing_angle}. The 0 degree case has hugely increased low energy absorption, due to the fact that observed photons must traverse the entire cloud before escaping. This would be similar to a spherical cloud that is surrounding the continuum source, see \cite{leahy93}. The iron edge depth decreases with increasing angle. The large increase in total continuum flux with increasing angle is caused by an interplay of two separate effects. The first being, at energies $5-50$ keV, the Compton scattering distribution is Thomson like, i.e. photons are as likely to scatter backward as forwards. The second due to the fact that most photons are scattered from the cloud from a very shallow depth. After a photon is scattered in the direction of the observer it is progressively less likely to be absorbed before escaping with increasing scattering angle. It can be seen in Figure \ref{fig:changing_angle} that, as energy increases, the disparity in flux reduces and reverses. Note the reversal in flux between the 60 and 150 degree cases around 60 keV, which is where the Compton scattering distribution becomes non-Thomson in nature and begins to favour forward scattering. The large increase in the flux for the 0 degree case is not caused by the same process due to the fact that the energy is too low for forward scattering to be favoured. However, at energies $>$ 10 keV absorption becomes increasingly unlikely, thus photons are scattering not just on the surface of the cloud but have the chance to pass through the entire length of the cloud, leading to  increased flux at energies $>$ 10 keV. It should be noted that as discussed in section \ref{the_mc_code} photons that do not scatter at all do not contribute to the spectrum.

\begin{figure}
\includegraphics[scale=0.7]{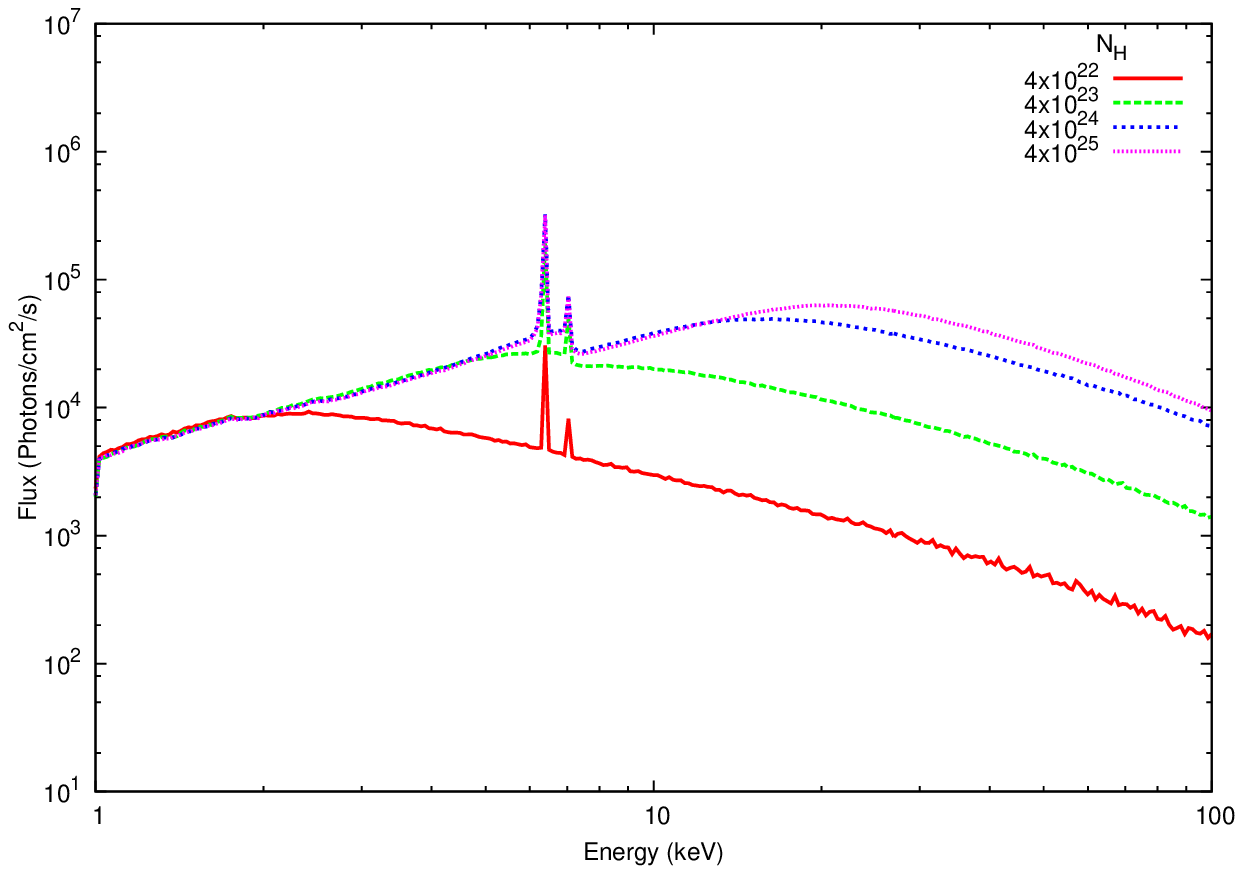}
 \caption{Spectra produced by the MC code for a 2 pc diameter, uniform density sphere positioned with a line of sight angle $\theta = 140^{\circ}$, and a photon index of $\Gamma = 2.0$. Which shows the changes in flux and continuum shape resulting from a changing $N_H$.}
  \label{fig:changing_nh_hiangle}
\end{figure}

\begin{figure}
\includegraphics[scale=0.7]{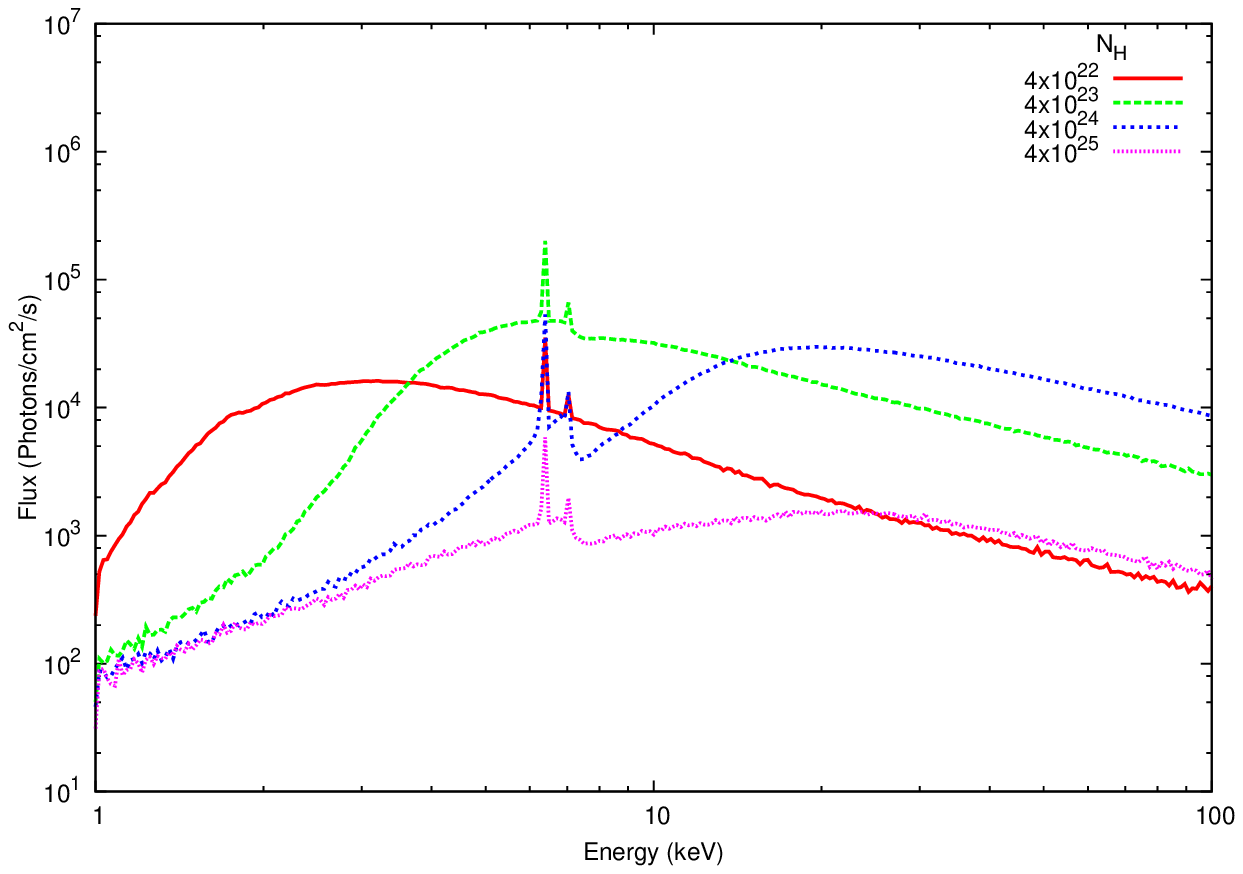}
 \caption{Spectra produced by the MC code for a 2 pc diameter, uniform density sphere positioned with a line of sight angle $\theta = 30^{\circ}$, and a photon index of $\Gamma = 2.0$. Which shows the changes in flux and continuum shape resulting from a changing $N_H$.}
  \label{fig:changing_nh_lowangle}
\end{figure}

In the case of varying column density $N_H$ (see Figures \ref{fig:changing_nh_hiangle} \& \ref{fig:changing_nh_lowangle}) we see that in the high angle case (Figure \ref{fig:changing_nh_hiangle}) low energy absorption is unaffected by increasing $N_H$ due to the fact that all of photons are back scattered from the cloud from a very shallow depth. The effect is far less pronounced in the low angle case (Figure \ref{fig:changing_nh_lowangle}). In the high angle case at low energies ($<$ 10 keV) differences in continuum shape and flux become very small after a critical $N_H$ is reached ($N_H > 5\times10^{23}$ cm$^{-2}$). Of note is the fact that in the low angle case (Figure \ref{fig:changing_nh_lowangle}) the high column density $N_H = 4\times10^{25}$ cm$^{-2}$ has very low flux in the high energy. We find that this is due to increased multiple scattering away from the observer, with the effect becoming more pronounced as multiple scattering increases.

\subsection{Iron Line Flux}

\begin{figure}
\includegraphics[scale=0.7]{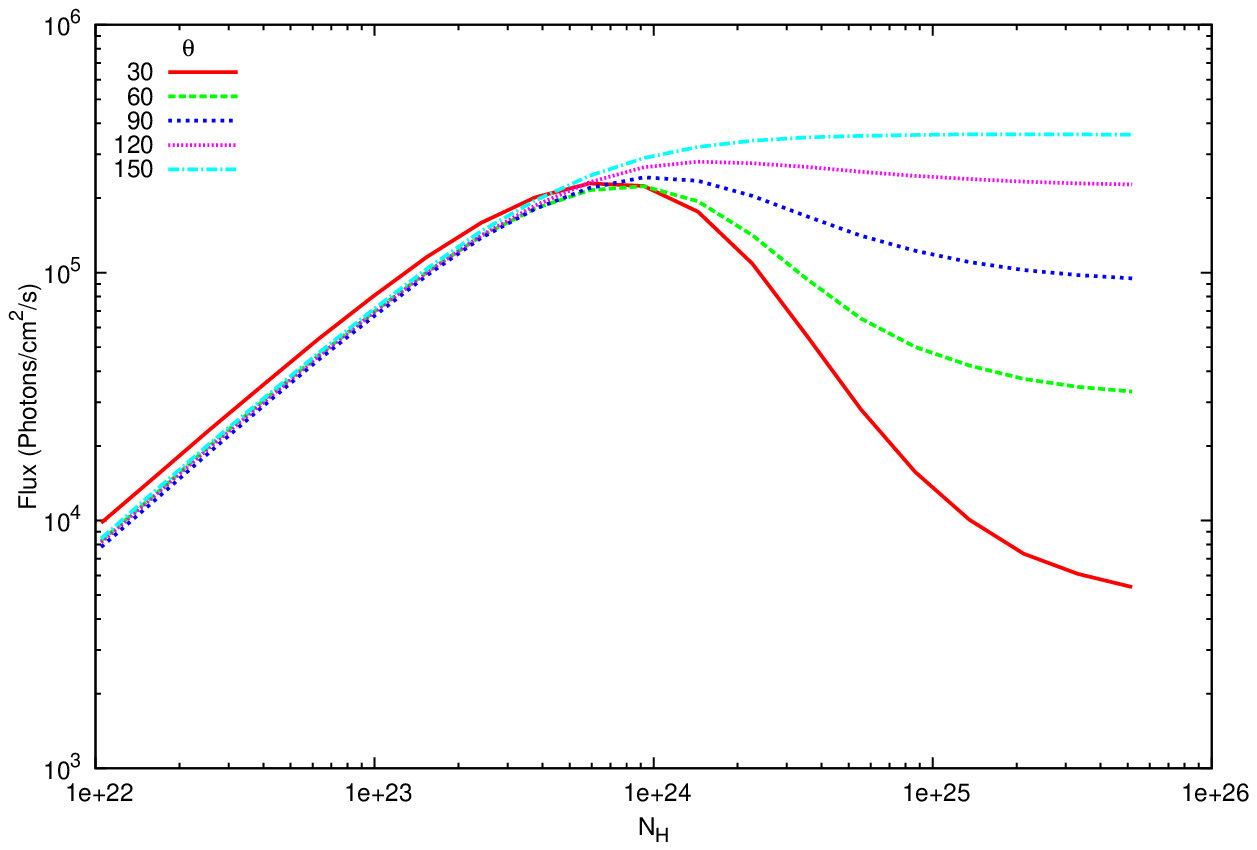}
\includegraphics[scale=0.7]{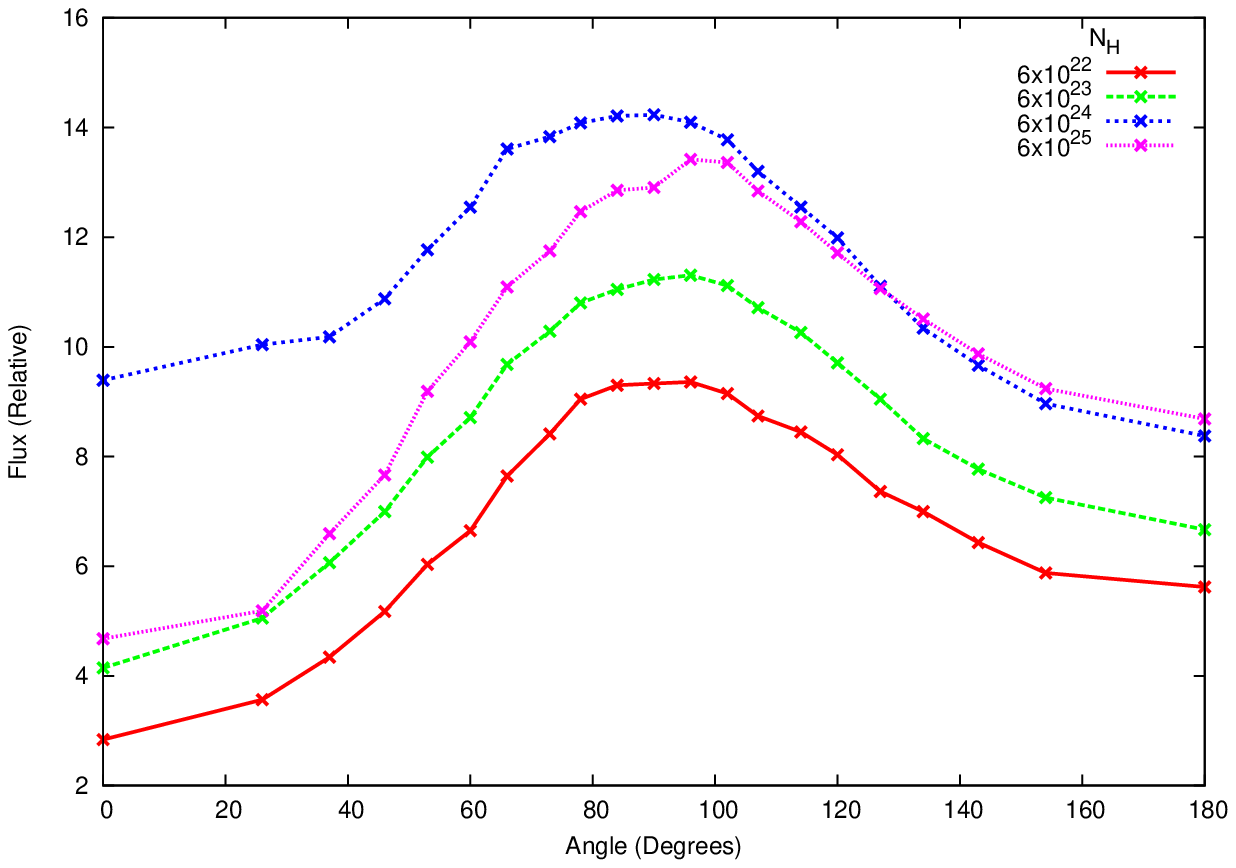}
\caption{\newline
		 Top panel: The dependence of the strength of the Fe $K_{\alpha}$ line versus varying $N_H$ for various angles.
		 \newline
		 Bottom panel: The dependence of the relative strength of the Fe $K_{\alpha}$ line for different $N_H$ values versus varying angle $\theta$.  
	    }
\label{fig:rel_flux}
\end{figure}

We show in Figure \ref{fig:rel_flux} how the total iron line flux varies with increasing column density $N_H$. There is an increase in flux up to an $N_H$ of ~$6\times10^{23}$ cm$^{-2}$ and then a gradual decrease in agreement with \cite{sunyaev98} who gave a maximum Thomson optical depth of $\approx$0.4 for maximum iron line flux. Interestingly there is a plateauing of the flux as angle increases, again due to the fact that we are only seeing the surface of the cloud. The relative $K_\alpha$ flux versus line of sight angle is also shown (bottom panel), peaking at $\theta = \frac{\pi}{2}$. The relative flux is the strength of the iron line above the continuum.

\subsection{Non-uniform density}
\begin{figure}
\centering
\includegraphics[scale=0.7]{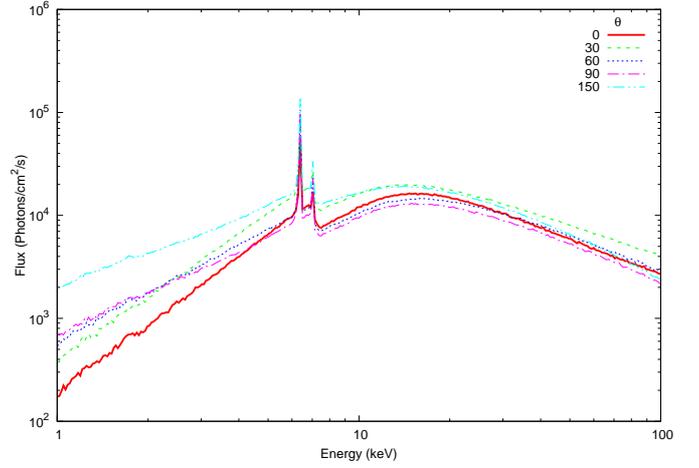}
 \caption{Similar comparison as that shown in Figure \ref{fig:changing_angle}, except for a variable density sphere. A Gaussian density profile, with a central density of $n_0 = 1.2\times10^5 cm^{-3}$ ($N_H = 2.5\times10^{24}$), $\sigma$ = 1.1 pc, and a step size of 0.1 pc. Showing the changes induced by a changing line of sight angle.}
  \label{fig:gaus_changing_angle}
\end{figure}

\begin{figure}
\includegraphics[scale=0.7]{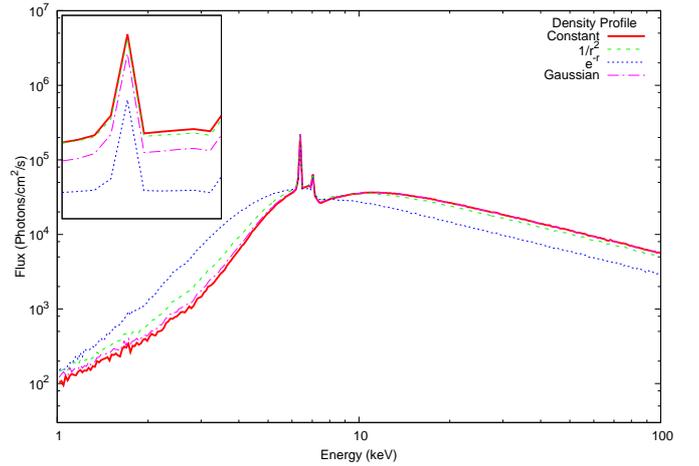}
 \caption{Comparison of the four utilized density profiles for a sphere of constant mass, radius is changing, the line of sight angle $\theta = 30^{\circ}$. $n_0 = 1.2\times10^5 cm^{-3}$ ($N_H = 2.5\times10^{24}$), the variable density cases have a step size $\kappa$ = 0.1 pc. The smaller figure in the top left showing the iron line has been normalised by incident flux.}
  \label{fig:profile_comparison}
\end{figure}

As discussed in section \ref{the_mc_code}, the code is able to model density profiles other than uniform. We calculate spectra for the $r^{-2}$, $e^{-r}$ and Gaussian density profiles. A step size of $\kappa = 0.1$ pc is used for this comparison, in a sphere of radius 2 pc. Figure \ref{fig:gaus_changing_angle} demonstrates that changes in line of sight angle are still readily apparent for the Gaussian case. Notably the point at which the 30 degree and 90 degree continuum flux reverse is a few keV lower than uniform density. This is due to surface scattering no longer being as dominant. Figure \ref{fig:profile_comparison} shows a comparison between density profiles for spheres of constant mass, with line of sight angle held constant, demonstrating the slight differences between density profiles. The Gaussian has a standard deviation of $\sigma$ = 1.1 pc.

\section{Observations and data reduction} \label{observations}
Sgr B2 was observed by \textit{Chandra} for 100 ks in July 2000 when it was very bright. We have analysed the corresponding dataset (ObsID 944) using \texttt{ciao} v 4.3. Event lists were cleaned for high background periods yielding effective exposures of respectively 97 ks. 

\begin{figure}
\includegraphics[scale=0.35]{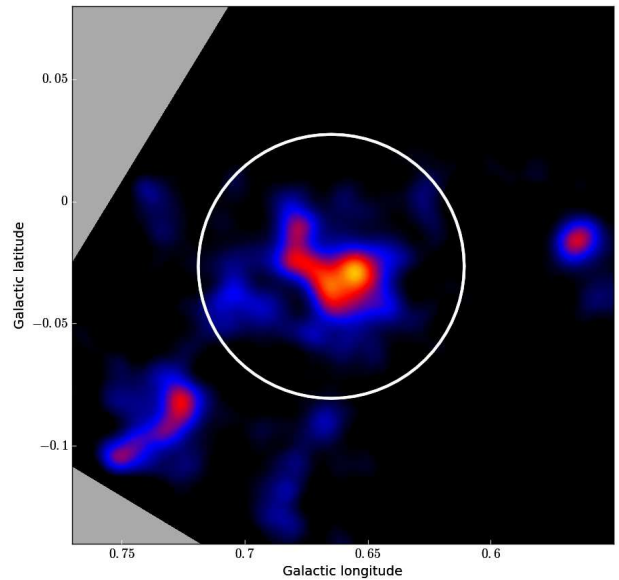}
 \caption{6.4 keV line flux image obtained with \textit{Chandra} in 2000. The bright core of Sgr B2 in the centre, surrounded by more diffuse emission. The \textit{Chandra} and \textit{XMM-Newton} spectra have been extracted from a region (white circle) of $3.2'$ radius centred on the core of Sgr B2.}
\label{fig:datasetlocation}
\end{figure}

Images were extracted in the energy ranges 3-6 keV and 6.3-6.5 keV, which contains most of the 6.4 keV Fe K$\alpha$ line. To produce point source free diffuse emission maps, we performed source detection in the 3-6 keV energy range using the \texttt{wavdetect} task and removed the regions surrounding the source and filled the hole using the count rate estimated locally taking into account Poisson random fluctuations with the task \texttt{dmfilth}. Exposure maps were then produced in each energy interval assuming an absorbed power law spectral shape to weight the different energy contributions to the total exposure. We used the absorption to the Galactic centre ($N_H = 7\times10^{22} cm^{-2}$) and a power law index of 2. The background was estimated using blank field data provided in the calibration database (CALDB) and normalized to the images using the high energy count-rate. The point source cleaned images were then corrected for the background and divided by the corresponding exposure. Finally, we smoothed to produce flux images of the diffuse emission, see Figure \ref{fig:datasetlocation}.

We have extracted the spectrum from a region of 3.2 arcmin radius centred on the peak of the column density at l= 0.66$^\circ$ and b= -0.03$^\circ$ \citep{protheroe08}. The spectral extraction region radius of 3.2 arcmin (7.5 pc at GC distance) was chosen because it provides a full coverage of the cloud with 99\% of the mass enclosed according to the density profile obtained by \cite{protheroe08}. It is also chosen to provide a relatively good match with the INTEGRAL IBIS/ISGRI PSF ($12^\prime$ FHWM) and to ensure a stable comparison between the soft X-rays and hard X-ray datasets. Spectral extraction and instrument responses were generated using the \texttt{specextract} task. Because of the intensity of the local astrophysical background and its variations in the field of view, we used blank sky observations provided by the CXC to estimate the background contribution in each region.

We analysed the \textit{XMM-Newton} data taken in September 2004 during a dedicated 50 ks exposure of the Sgr B2 molecular cloud (ObsID 0203930101). Because of significant contamination by solar flare events we did not use the EPIC/PN data. After cleaning the high background time intervals, the resulting exposure time with the EPIC/MOS cameras is 40 ks. We extracted the spectra from the EPIC/MOS instruments using the Extended Source Analysis Software (ESAS) \citep{snowden08} distributed with version 12.0.1 of the \textit{XMM-Newton} Science Analysis Software.

We produced calibrated and filtered event files with the tasks \textit{emchain} and \textit{mos-filter}, in order to exclude the time intervals affected by soft proton contamination.  The spectra were extracted from the same region as the \textit{Chandra} using the ESAS \textit{mos-spectra} scripts and were re-binned to have at least 30 counts in each bin to apply chi-square statistics.  The background was obtained from the filter wheel closed archival observations provided within the ESAS database. It provides an estimate of the quiescent component of the EPIC internal particle background.

In order to have a good constraint on the hard X-ray component of the spectrum, we have used the INTEGRAL data presented in \citet{terrier10}. They cover observations taken during the years 2003 and 2004, roughly contemporaneous with the \textit{XMM-Newton} data. We refer to this paper for details on data treatment, calibration and spectrum extraction. We recall that the INTEGRAL source was found perfectly coincident with the Sgr B2 core region visible in figure \ref{fig:datasetlocation}. The X-ray emission and the hard X-ray spectrum seen with Integral are therefore most likely due to the same phenomenon. A summary of the data used can be seen in Table \ref{tbl:obsdetails}.

\begin{table*}
\centering
\caption{Observation details}
\label{tbl:obsdetails}
\begin{tabular}{ccccc}
ObsID & Observatory & Instrument & Start & Duration (ks) \\
\hline
944 & ACIS-I & Chandra & 2000-03-29 09:43:32 & 100 \\
0203930101 & EPIC/MOS & XMM-Newton & 2004-09-04 02:52:56 & 50 \\
010300360010 - 02490026010 & IBIS/Isgri & INTEGRAL & 2003 & 7100 
\end{tabular}
\end{table*}

\section{Constraining properties of Sgr B2} \label{sgrb2}
The Xspec table models were created with the following free parameters, Line of sight angle $\theta$, Incident photon index $\Gamma$ and cloud column density $N_H$. Although variable before model creation, the cloud size, distance to source, distance to observer, polar angle, and elemental abundances are all fixed. The free parameter ranges are $\Gamma = 1.4 - 3.0$, $N_H = 1.02\times10^{22}  - 5.04\times10^{25}$ cm$^{-2}$, and $\theta = 0 - \pi$. The $\Gamma$ parameter has a step size of 0.1, the $N_H$ parameter has 20 steps that increase logarithmically from minimum to maximum, finally the Angle parameter is split into 21 steps evenly spaced in cosine. For the non-uniform density models, the $N_H$ parameter is actually in terms of the central density $\rho_0$.

We fit our data simultaneously with all parameters linked except normalisations, which are kept separate for the \textit{Chandra} data. We fit the data with the table model + a double \textit{apec} warm plasma model (to account for the galactic centre plasma emission) as follows \textbf{wabs(apec+apec+Model)}. The plasma temperatures are not well constrained by the model, but are necessary for a good quality fit. The thermal diffuse emissions in the GC are usually described by a two-component plasma model with typical temperatures on the order of 1 keV and 7 keV \citep{koyama89, tanaka00}. In the central regions, \cite{muno04} find temperature values of 0.7-0.9 keV for the cool component and 6-8 keV for the hot ones. Numerous studies with \textit{Suzaku} have shown the soft plasma to have roughly a temperature of 1 keV in many regions in the CMZ, while \cite{koyama07} found that the hot plasma is well described by a collisional plasma with temperature 6.5 keV and solar metallicity. In the Sgr B2 region, \cite{ryu09} used a model with 3 APECs. The first one is not absorbed and contribute below 2 keV and is thought to be local. We excluded this energy range and neglected this contribution. The second, the soft plasma temperature, was fitted and they found kT = 0.7 keV. The third, the hot plasma temperature, was fixed at 6.5 keV. We follow the same approach but we also fix the soft component temperature. Note that taking into account the absorption to the GC the main effect of a change of 0.7 - 1 keV above 2 keV is mainly in the relative intensities of the plasma lines. We thus fix the plasma temperatures to 1 keV and 6.5 keV respectively. We performed this fitting for both uniform and non-uniform density, namely Gaussian.

\begin{figure}
\includegraphics[scale=0.35]{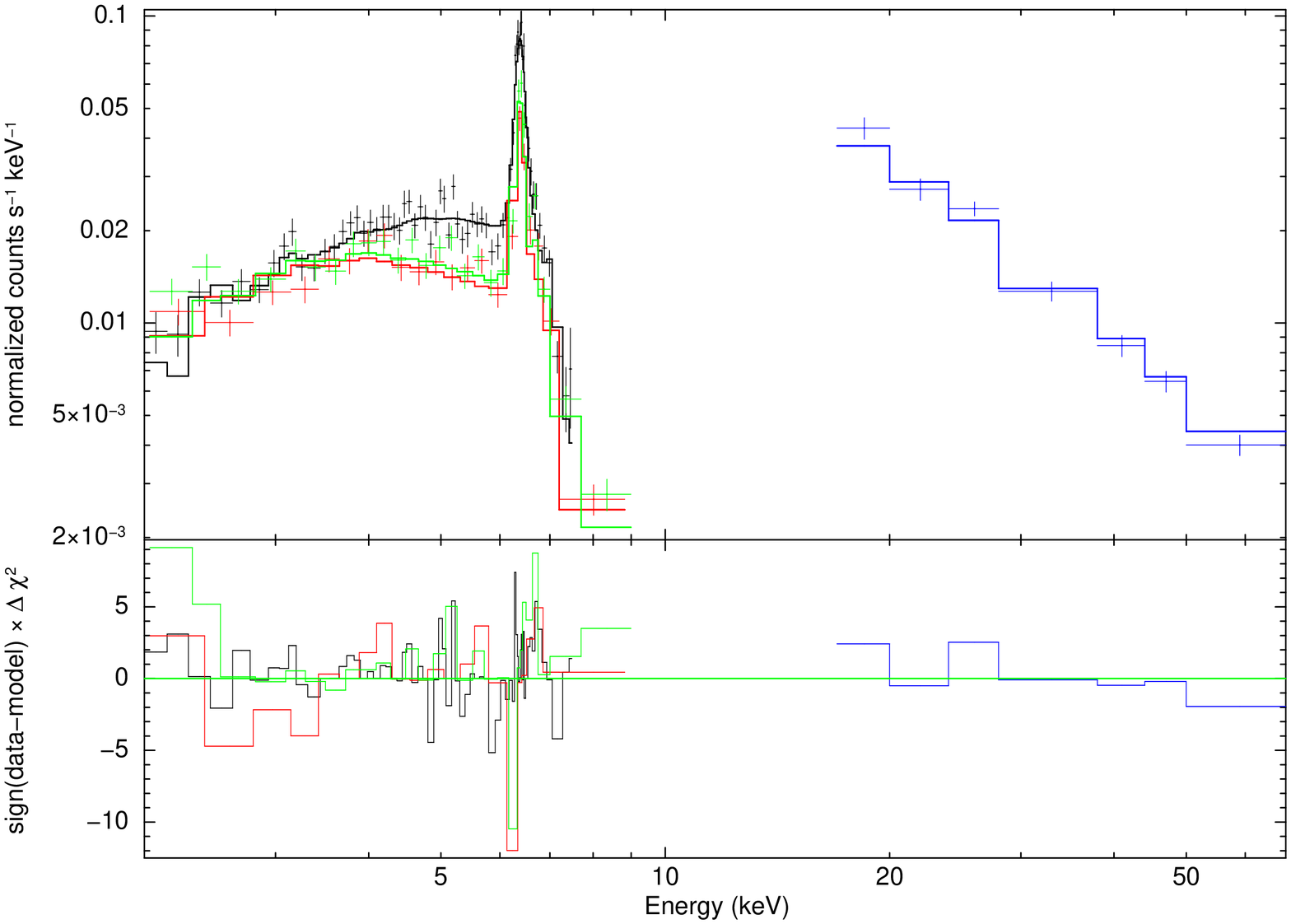}
 \caption{Plot of best simultaneous fit to \textit{Chandra} + \textit{XMM-Newton}/\textit{Integral} data for the uniform density table model. The black is \textit{Chandra} (2000), red and green are EPIC/MOS1 and EPIC/MOS2 (2004) respectively, and blue is Integral (2004). The normalizations of the two periods are independent but all other parameters are considered fixed across data sets.}
\label{fig:bestfitplot}
\end{figure}

The results of the uniform density fit can be seen in Figure \ref{fig:bestfitplot} and Table \ref{tbl:bestfittbl}. While the Gaussian fit can be seen in Figure \ref{fig:bestfitgausplot} and Table \ref{tbl:bestfittbl}.  Contour plots of both fits can be seen in Appendix \ref{appendixplots}.

For the uniform density the fitted angular position $\theta = 64_{-7}^{+8}$ of Sgr B2 implies a distance from Sgr $A^\star$ to Sgr B2 of $111^{+8}_{-6}$ pc assuming a projected distance of 100 pc. Taking a distance to Earth of ~8.4 kpc, we obtain an incident luminosity for both the \textit{Chandra} data and the XMM-Newton/Integral data. The \textit{Chandra} data giving $L_{2-10} = 1.57\times10^{39} erg/s$. The XMM-Newton/Integral data gives $L_{2-10} = 1.13\times10^{39} erg/s$ see Table \ref{tbl:fluxtbl}.

The Gaussian parameters were obtained from \cite{protheroe08}, we used a cloud radius of 12 pc and a standard deviation $\sigma$ = 2.75 pc. We used a step size $\kappa$ = 0.6 pc. The fitted angular position $\theta = 89_{-11}^{+10}$ of Sgr B2 implies that Sgr B2 lies at it's projected distance from Sgr $A^\star$ of 100 pc. For the luminosity the \textit{Chandra} data gives $L_{2-10} = 2.03\times10^{39} erg/s$. The XMM-Newton/Integral data gives $L_{2-10} = 1.48\times10^{39} erg/s$ again shown in Table \ref{tbl:fluxtbl}.

\begin{figure}
\includegraphics[scale=0.35]{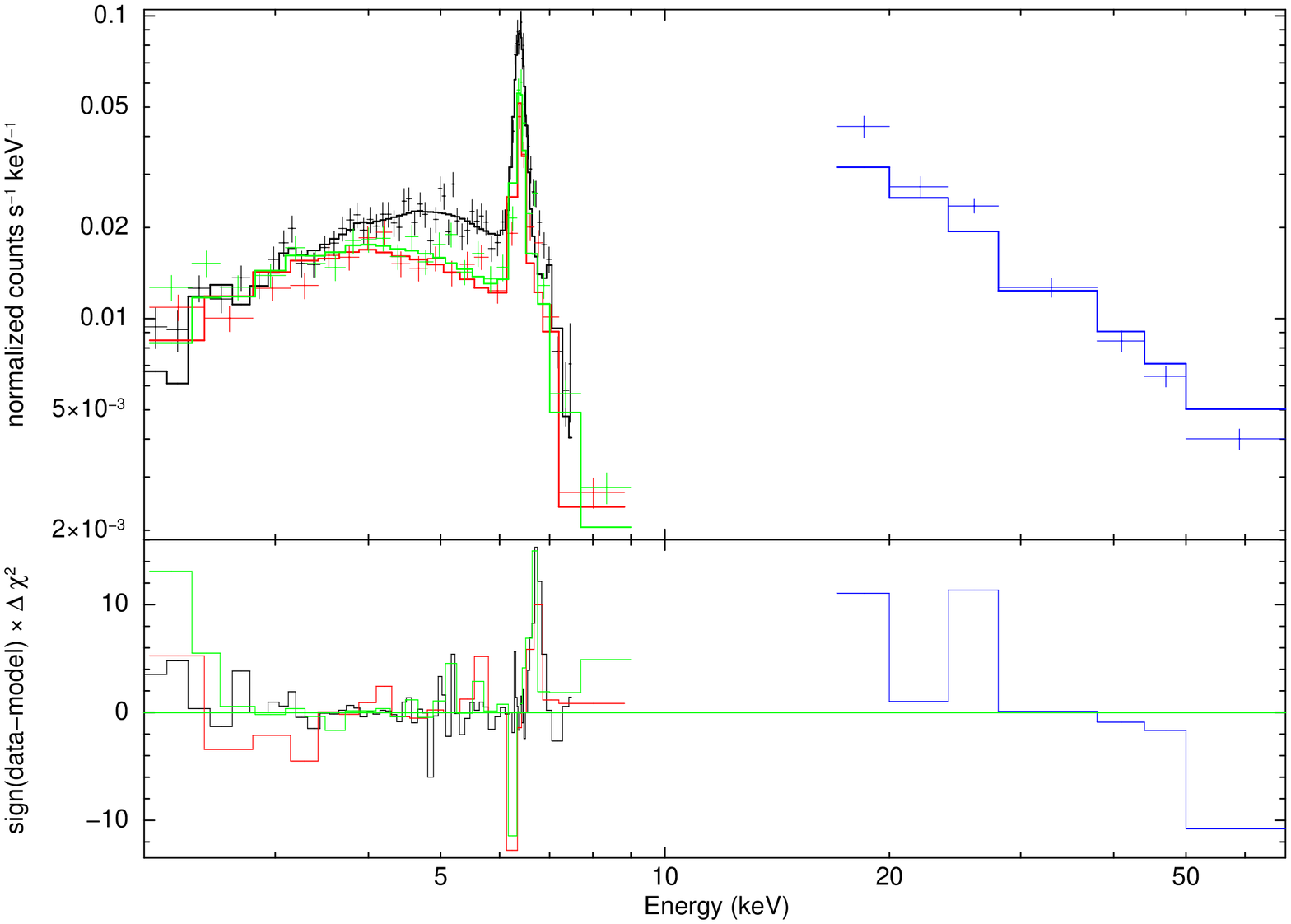}
 \caption{Plot of best simultaneous fit to \textit{Chandra} + \textit{XMM-Newton}/\textit{Integral} data for the Gaussian density table model. The black is \textit{Chandra} (2000), red and green are EPIC/MOS1 and EPIC/MOS2 (2004) respectively, and blue is Integral (2004). The normalizations of the two periods are independent but all other parameters are considered fixed across data sets.}
\label{fig:bestfitgausplot}
\end{figure}

\begin{table*}
\centering
\caption{Best fit parameters for uniform and Gaussian density table model to \textit{Chandra} + \textit{XMM-Newton}/\textit{Integral} data.}
\label{tbl:bestfittbl}
\begin{tabular}{ccccccc}
\hline
Density Profile & $\theta$ & $N_H$ & $\Gamma$ & wabs $N_H$ & kT 1 \& 2 & $\chi^2 (dof)$\\
~	    & & $(10^{24} cm^{-2})$ &  & $(10^{22} cm^{-2})$ & (keV) & \\
\hline\\
Uniform & $64_{-7}^{+8}$ & $2.1_{-0.4}^{+0.3}$ & $2.2_{-0.2}^{+0.1}$ & $6.3_{-0.9}^{+0.9}$ & 1 \& 6.5 & 731 (618)\\[0.3cm]
Gaussian & $89_{-11}^{+10}$ & $2.5_{-0.7}^{+0.4}$ & $1.8_{-0.1}^{+0.1}$ & $6.8_{-0.8}^{+0.8}$ & 1 \& 6.5 & 820 (618)\\[0.3cm]
\hline
\end{tabular}
\end{table*}

\begin{table*}
\centering
\caption{Flux's and associated incident luminosities for \textit{Chandra} and \textit{XMM-Newton}/\textit{Integral} data, obtained from both uniform and Gaussian density fits}
\label{tbl:fluxtbl}
\begin{tabular}{ccccc}
		    & \multicolumn{2}{c}{Uniform Density}                                                                    & \multicolumn{2}{c}{Gaussian Density}                                                                   \\
\hline
Observation & \begin{tabular}[c]{@{}c@{}}$Flux_{2-10}$\\ $(10^{-12} erg/cm^2/s)$\end{tabular} & \begin{tabular}[c]{@{}c@{}}$L_{2-10}$\\ $10^{39} (erg/s)$\end{tabular} & \begin{tabular}[c]{@{}c@{}}$Flux_{2-10}$\\ $(10^{-12} erg/cm^2/s)$\end{tabular} & \begin{tabular}[c]{@{}c@{}}$L_{2-10}$\\ $10^{39} (erg/s)$\end{tabular} \\
\hline\\
Chandra                     & $7.3_{-0.2}^{+0.1}$                                & $1.6_{-0.4}^{+0.2}$                               & $7.0_{-0.1}^{+3.2}$                                & $2.0_{-0.4}^{+1.9}$ \\[0.3cm]
XMM-Newton/Integral         & $5.3_{-0.1}^{+0.1}$                                & $1.1_{-0.2}^{+0.2}$                               & $5.1_{-0.2}^{+1.7}$                                & $1.5_{-0.2}^{+0.5}$\\[0.3cm]
\hline
\end{tabular}
\end{table*}

A point of particular note is that the hard X-ray data is fundamental in properly constraining the spectral index. In regards to the abundance, many authors give an iron abundance greater than solar for Sgr B2, for example $\approx 1.3$ \citep{terrier10} and $\approx 1.9$ times Solar \citep{murakami01, revnivtsev04}. We made several runs of the code with higher iron abundances and fit the data with a uniform density table model set to 1.9 times solar iron abundance. Although the fit is technically worse, $\chi^2_{red} = 1.33$ compared to $\chi^2_{red} = 1.18$, it does better reproduce the iron line. Although this does suggest the abundance is higher than solar levels, we found the fitted parameters to be quite similar with incident luminosity $L_{2-10}$ increasing by $\approx 8\% - 13\%$ due to the increase in absorption caused by the higher iron abundance. Considering the similar parameters and that at higher iron abundances most changes in the continuum come at energies higher than 9 keV, also given to the fact that a proper equivalent width analysis is not within the scope of this paper we decided to leave all abundances at solar levels.

\section{Discussion and Conclusion.}  \label{discussion}
In this paper we presented a new Monte Carlo code for simulating X-ray reflection spectra from molecular clouds. The code is capable of modelling clouds of varying density distributions and of varying geometries. We show how different input parameters will result in highly divergent output spectra and provide an analysis of the processes behind these changes. Using \textit{Chandra}, \textit{XMM-Newton} and \textit{INTEGRAL} observations we constrain several parameters of the giant molecular cloud Sgr B2 using Xspec table models: the photon index $\Gamma$ of the incident spectrum, the luminosity of the illuminating source, the $N_H$ of the reflecting cloud and the angular position of the cloud relative to the line of sight.

\begin{figure}
\includegraphics[angle =-90, scale=0.35]{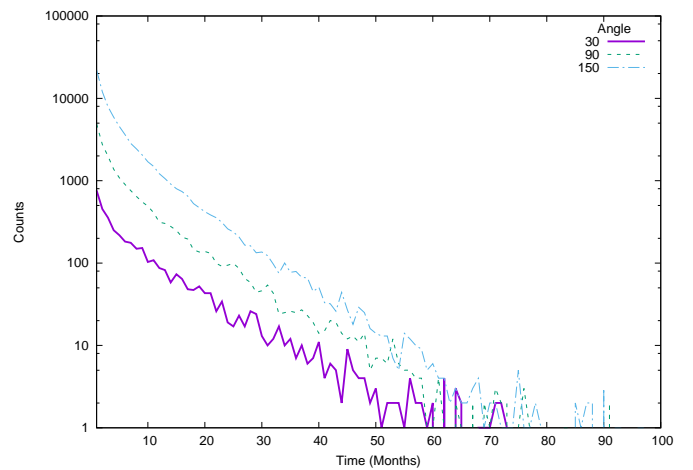}
 \caption{Light curve of escaping photons in the energy range 1 - 5 keV, for a Compton thick ($N_H = 4\times10^{24} cm^{-2}$) cloud. With various line of sight angle positions shown.}
\label{fig:lightcurve}
\end{figure}

More recent observations (such as \textit{NuSTAR}) of Sgr B2 could not be used directly with this model since it assumes the cloud is at or close to full illumination, and the hard X-ray flux has decreased by 40\% from 2003 to 2010 \citep{terrier10} with a time constant consistent with the light crossing time of the cloud. Given the cloud optical thickness, this is mostly due to multiply scattered photons \citep{sunyaev98}, but the main illuminating front has already started to leave the cloud. The relative brightness of the densest cores observed by \textit{NuSTAR} in hard X-rays \citep{zhang15} also supports this. Time dependent effects have therefore to be properly taken into account to model the spectra observed at later times. However, for the time interval between our observations we do not expect there to be any significant change in spectral shape. Figure \ref{fig:lightcurve} shows the light curve of observed photons in the 1-5 keV energy range for a Compton thick ($N_H = 4\times10^{24} cm^{-2}$) cloud for various line of sight positions. There is expected to be an extremely rapid decay in soft X-Ray flux after the cessation of the incident flare. The decay rate is best described by a power law, particularly in the first 20 months. Given the low characteristic time scale in the low energy, the observed time behaviour is likely close to that of the illuminating source. As there is an order of magnitude decay in the first year after flare cessation and we observe a rather marginal flux variation in 2004 compared to 2000, we can infer we are seeing a flare that lasted at least 4 years or close to it. In this case the spectral shape will not change between observations, assuming the incident spectrum remains the same.

Although the actual density profile of Sgr B2 is most certainly not uniform \citep{etxaluze13, jones11, protheroe08} it can be seen in Figure \ref{fig:profile_comparison} that differences between density profiles (excluding the exponential) are quite small, in fact it is unlikely that current X-ray data is capable of distinguishing the density profile. For this reason we presented both uniform Density and Gaussian density profile results. Both fit the data reasonably well. Importantly, they predict similar parameters. The Gaussian fit angle is slightly high $\approx90^{\circ}$ which is at odds with the literature but not completely unreasonable. However, the uniform density fit is technically superior with a $\chi^2_{red} = 1.18$ as opposed to $\chi^2_{red} = 1.32$ for the Gaussian. This poor Gaussian fit is perhaps due to the small centralized nature of the observations which precludes treating the observation as that of a cloud with a Gaussian distribution given the large size of the cloud. The column density predicted by both models ($N_{H_c} = 2.13\times10^{24} cm^{-2})$ and ($N_{H_G} = 2.5\times10^{24} cm^{-2}$) respectively, is higher than that obtained with X-ray observations previously, most likely due to the improper XRN models used. However, it is in good agreement with radio observations \citep{jones11, protheroe08}. The estimated incident luminosity on the order $10^{39}$ erg/s is in good agreement with the literature \citep{ponti13, zhang15}. There is a small decrease in Flux/Luminosity between the two periods, indicating a source luminosity decrease. The photon index found by both models $\Gamma_C = 2.16_{-0.16}^{+0.12}$ and $\Gamma_G = 1.83_{-0.11}^{+0.06}$ respectively, is also in good agreement with previous estimates \citep{revnivtsev04, terrier10, ponti13, zhang15}. This confirms that the illuminating source must have been a previous high activity period of Sgr $A^\star$. The fitted angle of the uniform density fit $\theta = 64_{-7}^{+8}$ degrees gives a Sgr $A^\star$ - Sgr B2 distance of $111^{+8}_{-6}$ pc (assuming a projected distance of $100$ pc) and places Sgr B2 $48^{+2}_{-1}$ pc closer to Earth than Sgr $A^\star$ in nominal agreement with \cite{kruijssen15} who give $\approx 38$ pc. Whereas the angle for the Gaussian fit $\theta = 89_{-11}^{+10}$ places the cloud at it's projected distance of 100 pc.

The ultimate aim of this fitting is to facilitate the analysis of GMC's in the Galactic Centre Region as reflectors of previous outbursts or periods of higher activity from Sgr $A^\star$. We make available these Xspec table models on Zenodo\footnote{\url{http://dx.doi.org/10.5281/zenodo.60229}}, we note that Iron abundance is not a parameter within the model, this was a choice to keep the number of fitting parameters low, however table models with other iron abundances are also available. We anticipate the use of this code in further work, fitting to other broadband X-Ray data (NuSTAR) from various GMC's. Helping to determine their characteristics and primarily their line of sight positions, thus offering improved constraints on the time delay and duration of the illuminating events.

\section{Acknowledgements}
The authors wish to acknowledge the DJEI/DES/SFI/HEA Irish Centre for High-End Computing (ICHEC) for the provision of computational facilities.\\
This work was partially supported by Ullyses Grant 2013.\\
The authors wish to acknowledge support from CNES.

\input{journals.tex}
\bibliography{gc}

\clearpage
\begin{appendix}

\section{Analytical calculations}
\subsection{Motivations}
In order to provide a comparison and a cross-check for the Monte-Carlo analysis, we perform a simple semi-analytic computation of the spectrum reflected by a cloud valid as long as multiple scattering is negligible. We first present the geometry used and the derivation of the spectrum neglecting multiple diffusion of the photons.
  
\subsection{Geometry of the problem}
We consider photons incident on a uniform spherical cloud of radius $R$ and density $n$. It is located at a distance $d$ of an illuminating source, which we will assume is far enough compared to $R$ so that the photons are all parallel when they reach the cloud surface. We have to compute the distribution of path lengths of the photons scattered within the cloud. We use a Cartesian frame $(x,y,z)$ centred on the cloud centre.

Let's assume the photons are incident on the cloud with a direction along the $z$ coordinate. 

A photon reaches the cloud at position $(x_0,y_0,z_0)$ before it propagates into it. After travelling a length $\lambda_{in}(z)$ it is scattered at point $M$ located at $(x_0,y_0,z)$. Its new direction makes an angle $\theta$ with the $z$ axis. The distance $\lambda_{out}(z,\theta)$ it will travel into the cloud before leaving it is

 $$\lambda_{out} = z\cos \theta -x_0\sin\theta + \sqrt{R^2 -\left(x_0\cos\theta^2+z\sin\theta\right)^2 -y^2}$$
 
\subsection{The scattered continuum} 
For simplicity, we neglect the effects of bound electrons and consider only Compton scattering. Its cross section is given by the Klein-Nishina formula $\frac{d\sigma_{KN}}{d\Omega}$.

The incident flux on the cloud is $F_0 = \frac{dN}{dE_0dt}\frac{1}{4\pi d^2}$, where $\frac{dN}{dE_0dt}$ is the number of photons emitted per unit time and unit energy. Now we consider a small volume element $dV = dxdydz$ located at $(x,y,z)$ inside the cloud. The incident photon flux it receives is given by $F_0 e^{-n\sigma_{tot}\lambda_{in}(z)}$.
where $\sigma_{tot} = \sigma_{KN}+\sigma_{abs}$ is the sum of the integrated Compton scattering and photoelectric absorption cross-sections. 

The flux scattered by the volume element is:
$$ \frac{dN}{d\Omega dt dE} = n \frac{d\sigma_{KN}}{d\Omega} dV F_0 e^{-n\sigma_{tot}\lambda_{in}(z)}$$

The fraction of the flux that will be absorbed or scattered over the distance $\lambda_{out}$ to the outer boundary of the spherical cloud is  $e^{-n \sigma_{tot} \lambda}$. Therefore we can write the flux scattered by the volume element $dM$ which is escaping the surface of the cloud to be:
$$ \frac{dN}{d\Omega dt} = n \frac{d\sigma_{KN}}{d\Omega} dV F_0 e^{-n\sigma_{tot}\left(\lambda_{in} + \lambda_{out}\right)} $$

Now we just have to integrate over the whole cloud to get the total flux scattered along the direction $\theta$.

So far, we have neglected the energy change of photons during the scattering process. The final energy is given by $E_s = E_0 / (1+\frac{E_0}{mc^2}(1-\cos\theta)$. Therefore, we have to distinguish the processes before and after the scattering: $\sigma_{tot}(E_0)$ and $\sigma_{tot}(E_s)$. We also need to include the Jacobian $\frac{dE_0}{dE_s}$.

The scattered flux is then:\\\\
$\frac{dN}{dE_s d\Omega dt} = n  \frac{d\sigma_{KN}}{d\Omega} \left[1-\frac{E_s}{m_ec^2}(1-\cos\theta)\right]^{-2} F(E_0) \times\\\\\\ 
 \times \int_{-R}^{R} dy \int_{-\sqrt{R^2-y^2}}^{\sqrt{R^2-y^2}} dx \int_{-\sqrt{R^2-x^2-y^2}}^{\sqrt{R^2-x^2-y^2}} dz\ \ e^{-n\sigma_{tot}(E_0)\left(\sqrt{R^2-x^2-y^2}-z\right)}  \times\\\\
  e^{-n\sigma_{tot}(E_s)\left(z\cos \theta -x\sin\theta + \sqrt{R^2-(x\cos\theta+z\sin\theta)^2)-y^2}\right)}$
\\\\which is valid up to $E_s \leq \frac{m_ec^2}{1-\cos\theta}$

We can factor out the $R$ constant and obtain a function of the column density $N_H$:\\\\
$\frac{dN}{dE_s d\Omega dt} =  N_H \frac{2 R^2}{\pi d^2}  \frac{d\sigma_{KN}}{d\Omega} \left[1-\frac{E_s}{m_ec^2}(1-\cos\theta)\right]^{-2} \frac{dN}{dE_0 dt}  \times\\\\\\  \times \int_{0}^{1} dw \int_{0}^{\sqrt{1-w^2}} dv \int_{0}^{\sqrt{1-v^2-w^2}} du\ \ e^{-\frac{1}{2}N_H\sigma_{tot}(E_0)\left(\sqrt{1-v^2-w^2}-u\right)}  \\\\\\
\times  e^{-\frac{1}{2}N_H\sigma_{tot}(E_s)\left(u\cos \theta -v\sin\theta + \sqrt{1-(v\cos\theta+ u\sin\theta)^2-w^2}\right)}$

\subsection{The fluorescence lines}
We now want to take into account the fluorescence lines, in particular the iron ones. At each point $(x,y,z)$, we compute the absorbed incident flux, integrating the incident spectrum with the energy dependent cross section for photo-electric absorption by iron, $\sigma_{Fe}$.

The flux of 6.4 keV photons emitted in a volume element $dV$ located at $(x,y,z)$ is:
$$ \frac{dN_{Fe}}{d\Omega dt} = \frac{1}{4\pi}Y_{Fe} n dV \int_{E_K} dE_0 \delta_{Fe}\sigma_{Fe}(E_0)  F(E_0) e^{-n\sigma_{tot}(E_0)\lambda_{in}}$$
with $\delta_{Fe}$ is the Fe abundance relative to hydrogen and $Y_{Fe}$ is the fluorescence yield.

Now the  flux escaping the cloud in the direction $\theta$ is:\\\\
$\frac{dN_{Fe}}{d\Omega dt} = \frac{1}{4\pi}Y_{Fe} n dV \int_{E_K} dE_0 \delta_{Fe}\sigma_{Fe}(E_0)  F(E_0) e^{-n\sigma_{tot}(E_0)\lambda_{in}} \times\\ e^{-n\sigma_{tot}(E_{Fe})\left(\lambda_{out}\right)} $\\
To get the full spectrum, one has to integrate over the cloud.

\section{Contour Plots}\label{appendixplots}
\begin{figure*}
\includegraphics[scale=0.35, angle=-90]{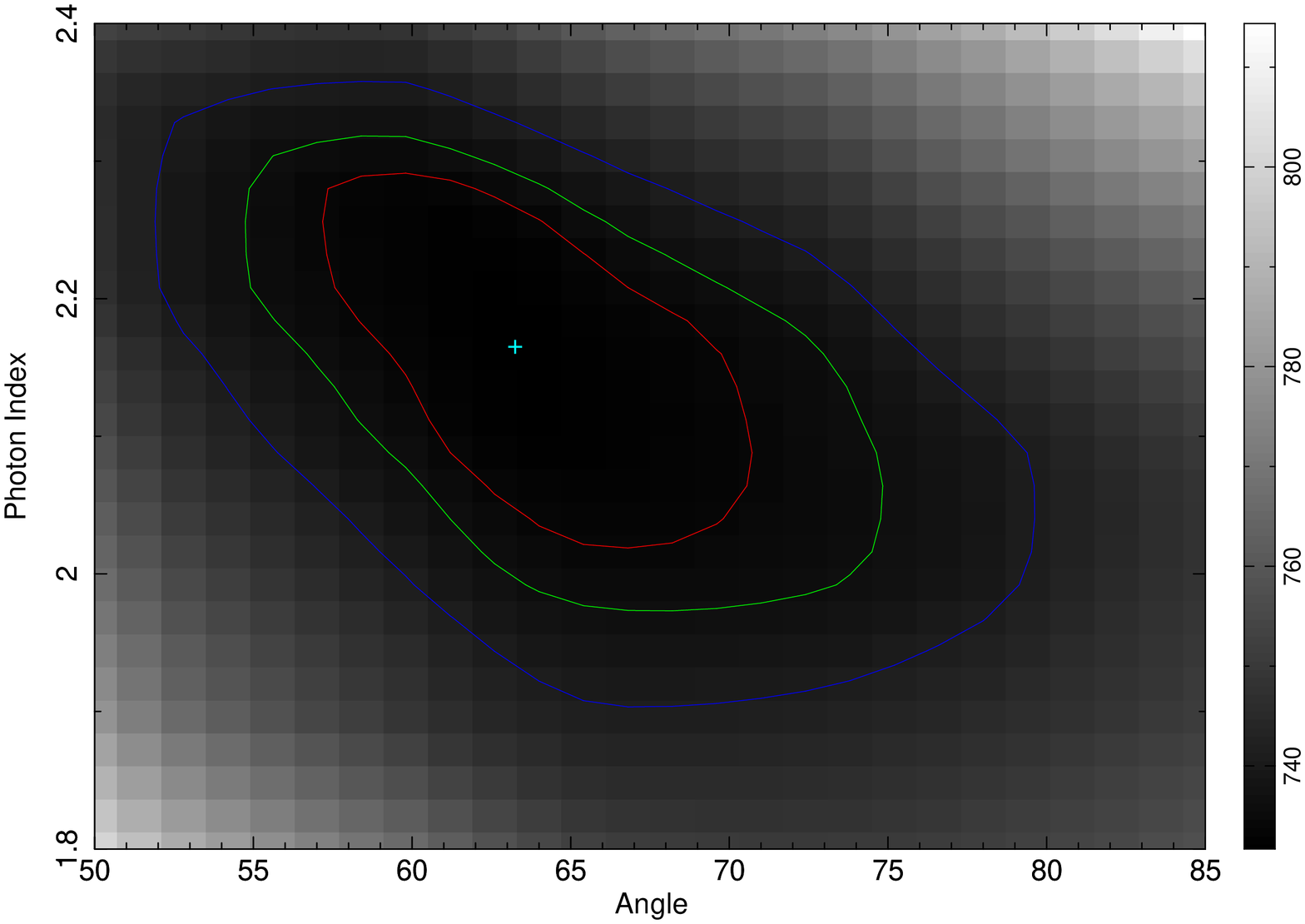}
 \caption{Contour plot of best simultaneous uniform density fit to \textit{Chandra} + \textit{XMM-Newton}/\textit{Integral} data. Angle -vs- Photon index.}
\label{fig:bestfitcontourplot1}
\end{figure*}

\begin{figure*}
\includegraphics[scale=0.35, angle=-90]{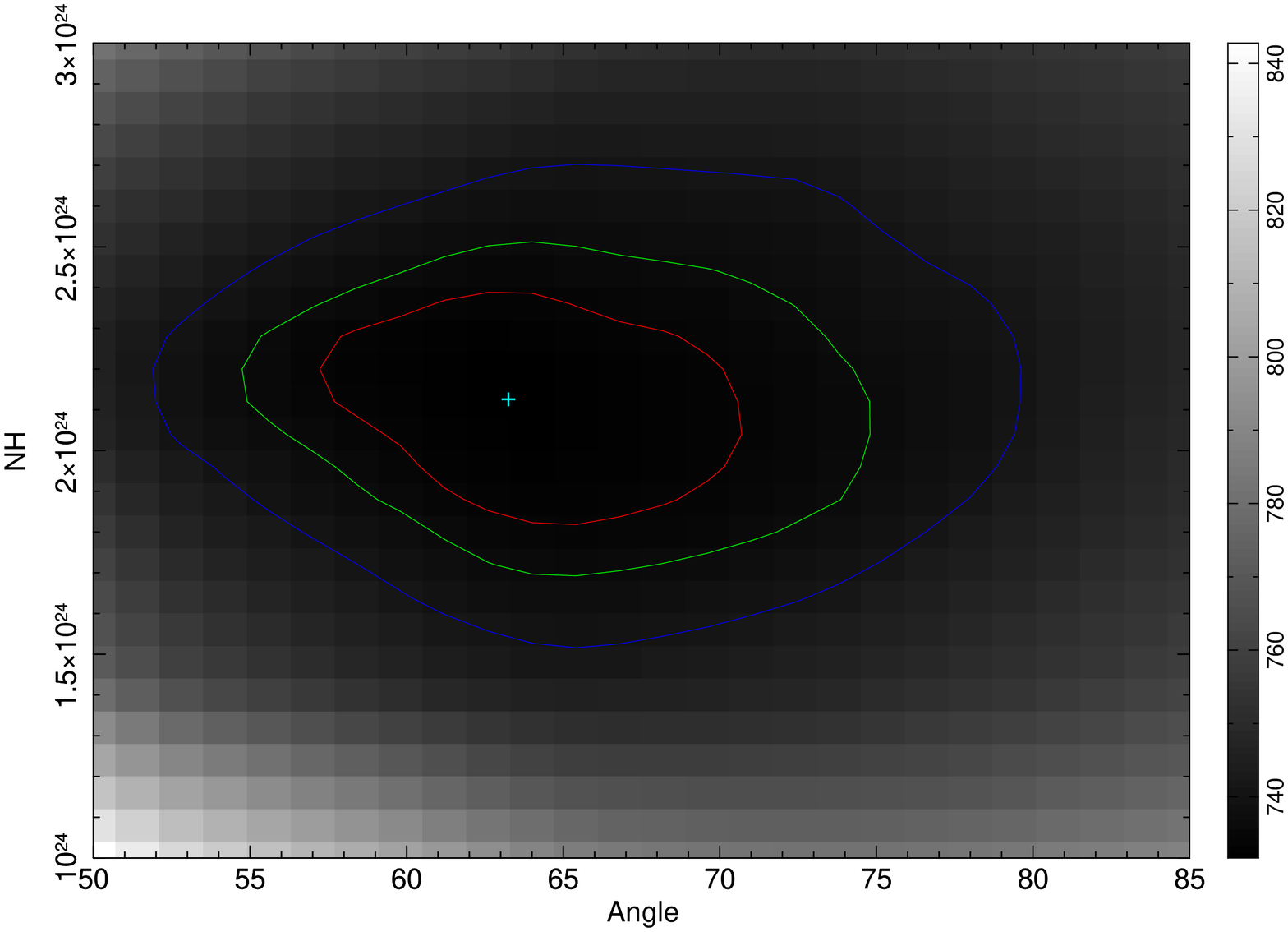}
 \caption{Contour plot of best simultaneous uniform density fit to \textit{Chandra} + \textit{XMM-Newton}/\textit{Integral} data. Angle -vs- $N_H$.}
\label{fig:bestfitcontourplot2}
\end{figure*}

\begin{figure*}
\includegraphics[scale=0.35, angle=-90]{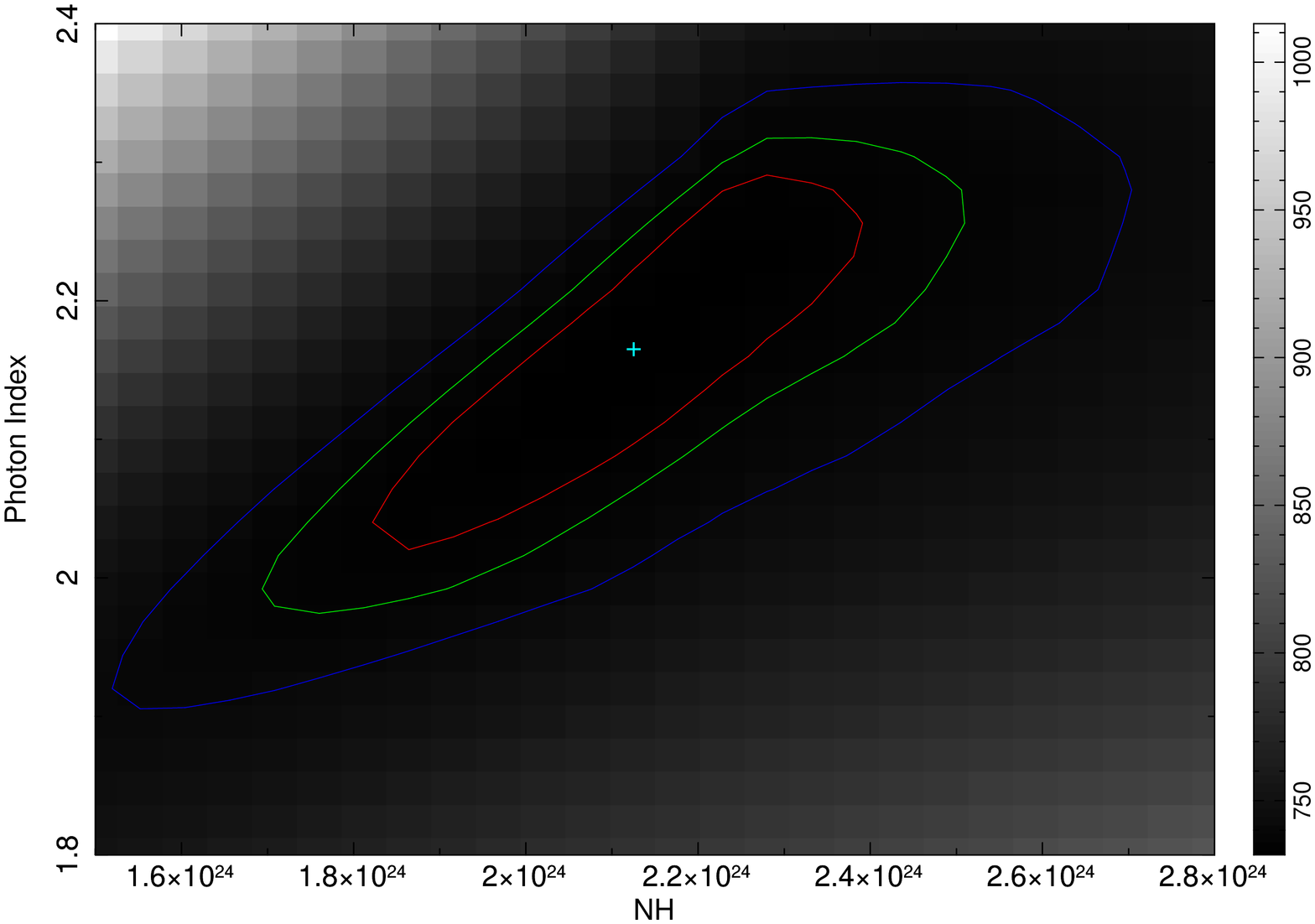}
 \caption{Contour plot of best simultaneous uniform density fit to \textit{Chandra} + \textit{XMM-Newton}/\textit{Integral} data. $N_H$ -vs- Photon index.}
\label{fig:bestfitcontourplot3}
\end{figure*}

Figures \ref{fig:bestfitcontourplot1} to \ref{fig:bestfitcontourplot3} are contour plots of the fit presented in section \ref{sgrb2} for the three fit parameters. The point marks the location of the best fit, the lines represent 1$\sigma$(red), 2$\sigma$(green) and 3$\sigma$(blue) inner to outer respectively. The shading is the $\chi^2$ value of the fit at that point based on the scale on the right side.

\begin{figure*}
\includegraphics[scale=0.35, angle=-90]{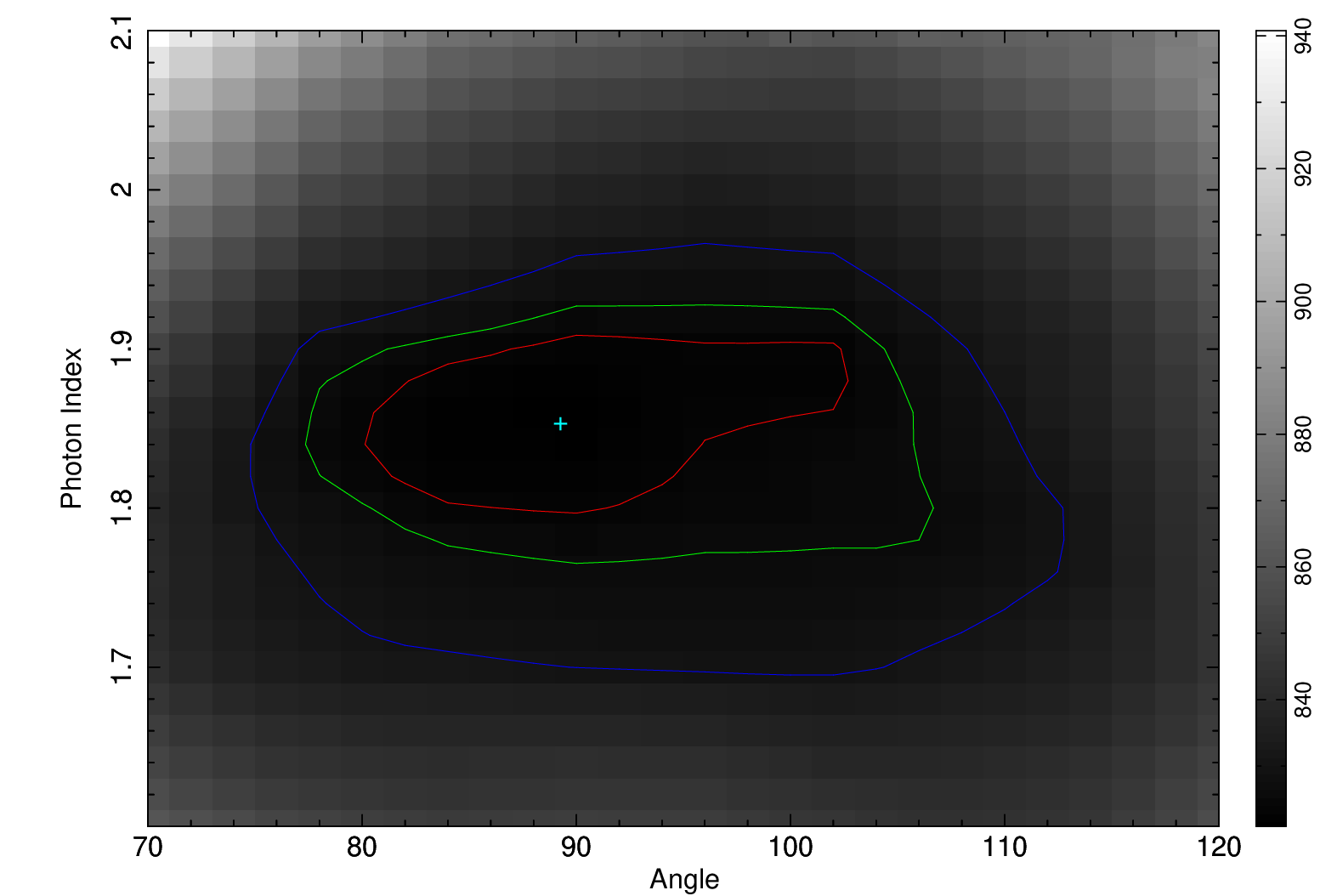}
 \caption{Contour plot of best simultaneous Gaussian density fit to \textit{Chandra} + \textit{XMM-Newton}/\textit{Integral} data. Angle -vs- Photon index.}
\label{fig:bestfitcontourplot4}
\end{figure*}

\begin{figure*}
\includegraphics[scale=0.35, angle=-90]{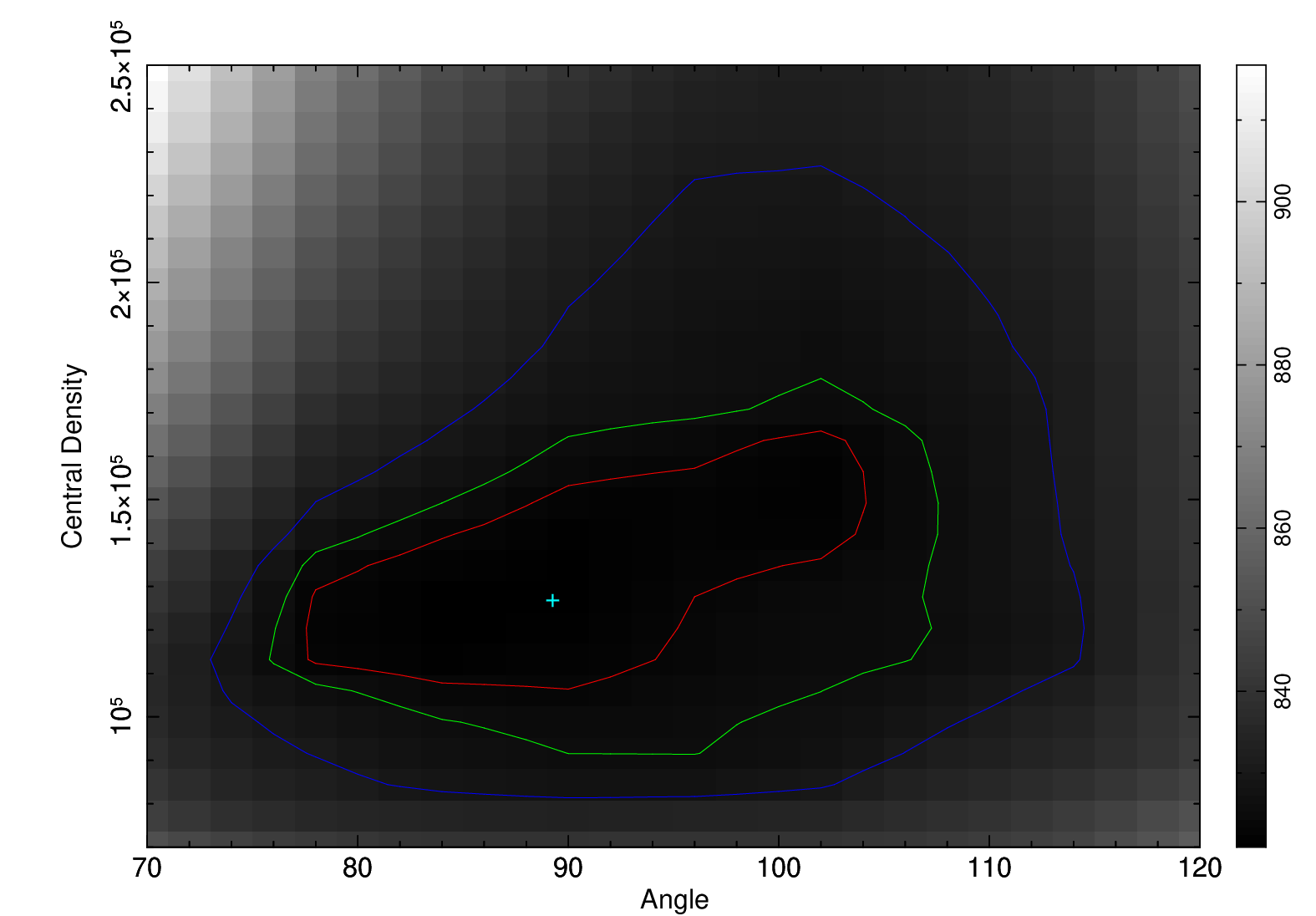}
 \caption{Contour plot of best simultaneous Gaussian density fit to \textit{Chandra} + \textit{XMM-Newton}/\textit{Integral} data. Angle -vs- Central Density.}
\label{fig:bestfitcontourplot5}
\end{figure*}

\begin{figure*}
\includegraphics[scale=0.35, angle=-90]{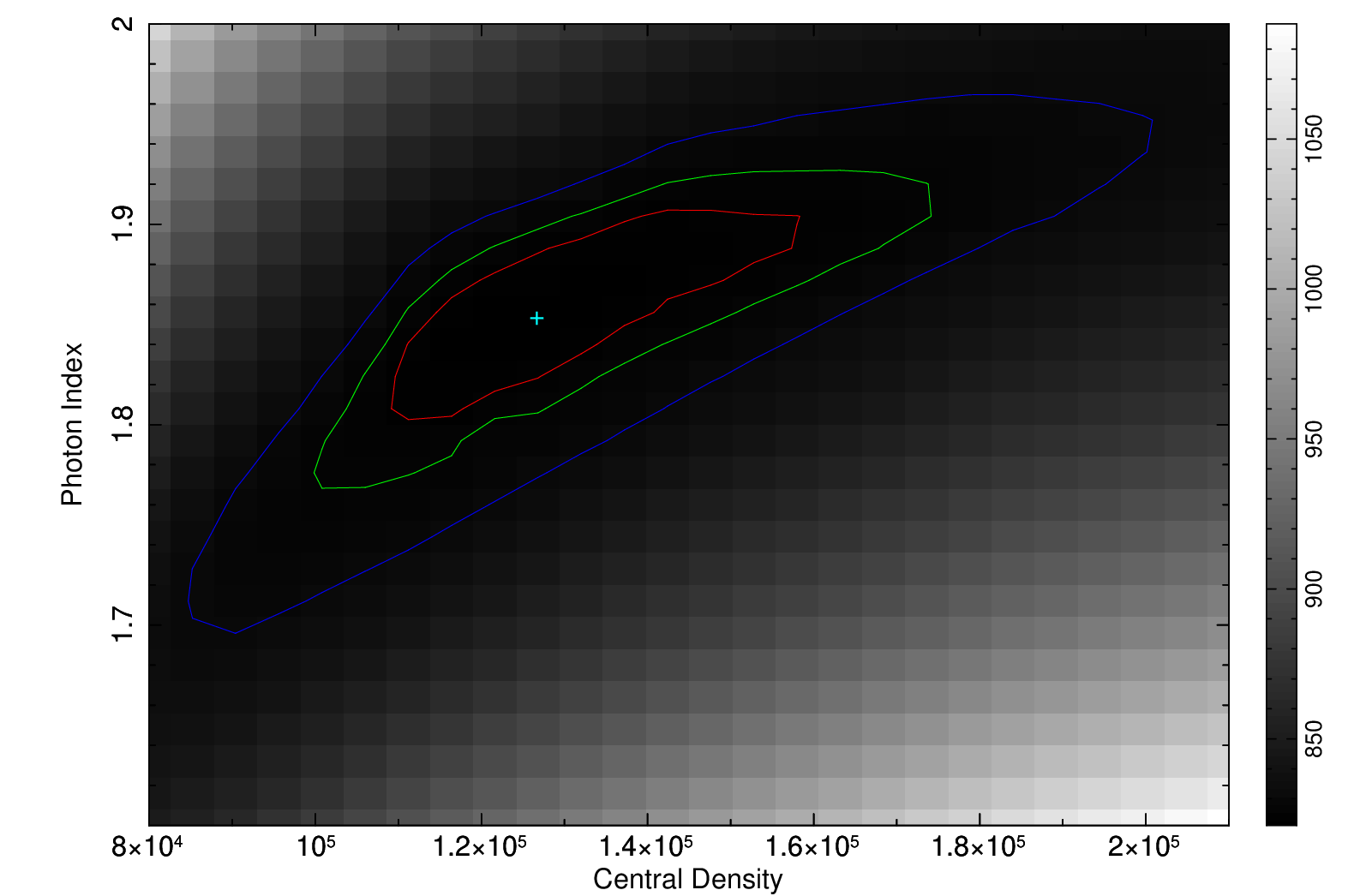}
 \caption{Contour plot of best simultaneous Gaussian density fit to \textit{Chandra} + \textit{XMM-Newton}/\textit{Integral} data. Central Density -vs- Photon index.}
\label{fig:bestfitcontourplot6}
\end{figure*}

\end{appendix}
\label{lastpage}

\end{document}

%% file: journals.tex
\def\aj{AJ}%
\def\actaa{Acta Astron.}%
\def\araa{ARA\&A}%
\def\apj{ApJ}%
\def\apjl{ApJ}%
\def\apjs{ApJS}%
\def\ao{Appl.~Opt.}%
\def\apss{Ap\&SS}%
\def\aap{A\&A}%
\def\aapr{A\&A~Rev.}%
\def\aaps{A\&AS}%
\def\azh{AZh}%
\def\baas{BAAS}%
\def\bac{Bull. astr. Inst. Czechosl.}%
\def\caa{Chinese Astron. Astrophys.}%
\def\cjaa{Chinese J. Astron. Astrophys.}%
\def\icarus{Icarus}%
\def\jcap{J. Cosmology Astropart. Phys.}%
\def\jrasc{JRASC}%
\def\mnras{MNRAS}%
\def\memras{MmRAS}%
\def\na{New A}%
\def\nar{New A Rev.}%
\def\pasa{PASA}%
\def\pra{Phys.~Rev.~A}%
\def\prb{Phys.~Rev.~B}%
\def\prc{Phys.~Rev.~C}%
\def\prd{Phys.~Rev.~D}%
\def\pre{Phys.~Rev.~E}%
\def\prl{Phys.~Rev.~Lett.}%
\def\pasp{PASP}%
\def\pasj{PASJ}%
\def\qjras{QJRAS}%
\def\rmxaa{Rev. Mexicana Astron. Astrofis.}%
\def\skytel{S\&T}%
\def\solphys{Sol.~Phys.}%
\def\sovast{Soviet~Ast.}%
\def\ssr{Space~Sci.~Rev.}%
\def\zap{ZAp}%
\def\nat{Nature}%
\def\iaucirc{IAU~Circ.}%
\def\aplett{Astrophys.~Lett.}%
\def\apspr{Astrophys.~Space~Phys.~Res.}%
\def\bain{Bull.~Astron.~Inst.~Netherlands}%
\def\fcp{Fund.~Cosmic~Phys.}%
\def\gca{Geochim.~Cosmochim.~Acta}%
\def\grl{Geophys.~Res.~Lett.}%
\def\jcp{J.~Chem.~Phys.}%
\def\jgr{J.~Geophys.~Res.}%
\def\jqsrt{J.~Quant.~Spec.~Radiat.~Transf.}%
\def\memsai{Mem.~Soc.~Astron.~Italiana}%
\def\nphysa{Nucl.~Phys.~A}%
\def\physrep{Phys.~Rep.}%
\def\physscr{Phys.~Scr}%
\def\planss{Planet.~Space~Sci.}%
\def\procspie{Proc.~SPIE}%
\let\astap=\aap
\let\apjlett=\apjl
\let\apjsupp=\apjs
\let\applopt=\ao